\documentclass[10pt,aps,prd,reprint,superscriptaddress,nofootinbib,preprintnumbers]{revtex4-1}

\usepackage[utf8]{inputenc}
\usepackage{amsmath}
\usepackage{amssymb}
\usepackage{graphicx}
\usepackage{multirow}
\usepackage[colorlinks,pdfusetitle]{hyperref}
\hypersetup{colorlinks=true,allcolors=[rgb]{1,0.56,0}}

\newcommand{\orcid}[1]{\href{https://orcid.org/#1}{#1}}

\DeclareMathOperator{\ci}{\text{i}}

\begin{document}

\title{A Survey of Neutrino Flavor Predictions and the Neutrinoless Double Beta Decay Funnel}

\author{Peter B.~Denton}
\email{peterbd1@gmail.com}
\thanks{\orcid{0000-0002-5209-872X}}
\affiliation{High Energy Theory Group, Physics Department, Brookhaven National Laboratory, Upton, NY 11973, USA}

\author{Julia Gehrlein}
\email{julia.gehrlein@colostate.edu}
\thanks{\orcid{0000-0002-1235-0505}}
\affiliation{Physics Department, Colorado State University, Fort Collins,   CO 80523, USA}
\affiliation{Theoretical Physics Department, CERN, 1 Esplanade des Particules, 1211 Geneva 23, Switzerland}
\affiliation{High Energy Theory Group, Physics Department, Brookhaven National Laboratory, Upton, NY 11973, USA}

\preprint{CERN-TH-2023-160}
\preprint{CETUP-2023-006}

\begin{abstract}
The neutrinoless double beta decay experimental effort continues to make tremendous progress with hopes of covering the inverted neutrino mass hierarchy in coming years and pushing from the quasi-degenerate hierarchy into the normal hierarchy.
As neutrino oscillation data is starting to suggest that the mass ordering may be normal, we may well be faced with staring down the funnel of death: a region of parameter space in the normal ordering where -- for a particular cancellation among the absolute neutrino mass scale, the Majorana phases, and the oscillation parameters -- the neutrinoless double beta decay rate may be vanishingly small.
To answer the question of whether this region of parameter space is theoretically preferred, we survey five broad categories of flavor model structures which make various different predictions for parameters relevant for neutrinoless double beta decay to determine how likely it is that the rate may be in this funnel region.
We find that a non-negligible fraction of predictions surveyed are at least partially in the funnel region.
Our results can guide model builders and experimentalists alike in focusing their efforts on theoretically motivated regions of parameter space.
\end{abstract}

\date{March 18, 2024}

\maketitle

\section{Introduction}
Neutrino oscillations \cite{Super-Kamiokande:1998kpq,SNO:2001kpb,SNO:2002tuh,ParticleDataGroup:2022pth} provide one of the few strong motivations for physics beyond the Standard Model (SM) as it requires at least two massive, active neutrinos.
Oscillation experiments, however, tell us about neither the absolute neutrino mass scale nor about the nature of the neutrino mass: Dirac vs Majorana.
One possible means of probing the latter question is to consider lepton number violating processes which provide a clear and striking signature that neutrinos are Majorana particles \cite{Schechter:1981bd}.
The most experimentally promising lepton number violating process is neutrinoless double beta decay (0$\nu\beta\beta$) which is the transition of a nucleus with (A, Z) atomic numbers to (A, Z + 2), accompanied by the emission of two electrons, but without the emission of two anti-neutrinos \cite{Furry:1939qr}.
The observation of neutrino oscillations has already demonstrated that the lepton number of individual flavors is not conserved; 0$\nu\beta\beta$ could go one step further marking the first observation that total lepton number is not a conserved symmetry of Nature either\footnote{Note that the converse need not be true.
That is, the non-observation of 0$\nu\beta\beta$, e.g.~if the atmospheric mass ordering was found to be inverted, does not guarantee that neutrinos are Dirac neutrinos.
One such scenario is pseudo-Dirac neutrinos \cite{Beacom:2003eu,Wolfenstein:1981kw,Bilenky:1983wt,Petcov:1982ya} where neutrinos are actually Majorana neutrinos but $|m_{\beta\beta}|$ may be quite small even in the inverted ordering.}.
This process is experimentally challenging to measure (for reviews see \cite{Vergados:2016hso, DellOro:2016tmg, Dolinski:2019nrj,Agostini:2022zub,Cirigliano:2022oqy}); however, experiments continue to make tremendous progress covering more and more parameter space pushing into the region suggested by neutrino oscillations.
Indeed, thanks to the progress in neutrino oscillation experiments, all neutrino mixing angles and mass splittings are now measured to a good accuracy \cite{Denton:2022een} which allows for improved predictions of the theoretically allowed regions of parameter space for $0\nu\beta\beta$ experiments. 

Somewhat surprisingly, the observed leptonic mixing pattern seems to be in considerable contrast to the quark mixing matrix, a difference that could imply a nontrivial connection between the two sectors.
Many models attempting to make sense of the so-called ``flavor puzzle" of the SM make predictions for the mixing parameters, including the so-called Majorana phases and absolute neutrino mass scale, which can be compared with experimental data. These models provide experimental targets for a wide variety of neutrino experiments and can be used to plan experimental stages or requested benchmark sensitivities \cite{IntensityFrontierNeutrinoWorkingGroup:2013sdv,Gehrlein:2022nss}.
Making precise predictions is challenging, however, due to the very large number of flavor models considered in the literature that still provide acceptable fits to existing neutrino (and possibly quark) data as they make a wide variety of predictions for the remaining neutrino parameters.

In this paper we will focus on the predictions related to flavor models applied to the neutrinoless double beta decay rate observable, motivated by the predicivity of flavor models for several parameters which enter this observable.
Taking the exchange of three light Majorana neutrinos as the dominant contribution to 0$\nu\beta\beta$, the predicted ranges for the particle physics observable $|m_{\beta\beta}|$ depend critically on the neutrino mass ordering which remains undetermined by oscillation data: normal (NO) with $m_1<m_2<m_3$ or inverted (IO) with $m_3<m_1<m_2$\footnote{We define the neutrino mass eigenstates in the usual way with decreasing amount of $\nu_e$ fraction: $|U_{e1}|>|U_{e2}|>|U_{e3}|$, see e.g.~\cite{Denton:2020exu}.}.
Of particular interest is the region in the NO which leads to immeasurably small rates of 0$\nu\beta\beta$ due to a precise cancellation among the absolute neutrino mass scale, the Majorana phases, and the oscillation parameters\footnote{While this region is essentially impossible to probe experimentally, it does have the advantage that if it was known that $|m_{\beta\beta}|$ were in the funnel, then we would have good knowledge of both Majorana phases \cite{Ge:2016tfx,Ge:2019ldu}, something that is otherwise essentially impossible.}; this region is known as the funnel and is often quantified as values of $|m_{\beta\beta}|<1$ meV for concreteness.
In this manuscript we will study this region of parameter space from a theoretical point of view in the context of a wide range of flavor models taking a bottom up approach (see \cite{Plentinger:2006nb,Jenkins:2008ex} for earlier studies before $\theta_{13}$ was measured).

We aim to provide a comprehensive study of viable categories of conceivable predictions phenomenologically related to the structure of existing flavor models discussed in the literature which make predictions for $|m_{\beta\beta}|$ and determine the fractions of predicted parameter space which fall into the funnel region within the constraints of the latest neutrino oscillation data.
That is, we are investigating whether or not categories of flavor models that have been studied in the literature containing any conceivable models prefer to be in the funnel.
Even though a particular focus of our work is the funnel region  we will also present a global overview of the preferred regions of parameter space in these phenomenological categories of models to demonstrate the existence and location of theoretically motivated regions of observables.
The advantage of doing such a study before experimental limits reach the normal hierarchy is to understand what ranges of observables models predict \emph{before} the measurements are made. This can guide future work both from the experimental and theoretical sides as we identify preferred regions of parameter space which can serve as targets to focus experimental efforts on. From the theoretical side we give a detailed overview of different categories of flavor models which make predictions for $0\nu\beta\beta$, assess their validity by comparing their predictions to current knowledge of the mixing parameters and bounds on the absolute mass scale, and calculate their preferred regions of parameter space. 
Our work can thereby provide important guidance for future model building work.
We focus on the low scale application of these predictions so our results do not depend on any details of neutrino mass generation; therefore, we do not include potential renormalization group effects on the running of the parameters that may be present in some scenarios.

This paper is organized as follows: we will start with a short introduction to 0$\nu\beta\beta$ in sec.~\ref{sec:0nubb}, then we explain and discuss the categories of predictions we consider including our results in secs.~\ref{sec:results}, \ref{sec:discussion}, and conclude in sec.~\ref{sec:conclusions}.

\section{Neutrinoless double beta decay review}
\label{sec:0nubb}
We start with a short review about neutrinoless double beta decay. 
We make the oft-used assumption that the dominant contribution to 0$\nu\beta\beta$ arrives from the exchange of three light ($m_\nu\lesssim$ 100 MeV \cite{Blennow:2010th}) Majorana neutrinos; see \cite{Deppisch:2012nb,Deppisch:2015qwa,Peng:2015haa,Fuks:2020zbm,Mitra:2011qr,Lopez-Pavon:2012yda,Blennow:2010th,Tello:2010am,Li:2020flq} for other new physics scenarios which give rise to 0$\nu\beta\beta$.
The observable in neutrinoless double beta decay is the decay half-life which is a function of various physics parameters,
\begin{align}
(T_{1/2}^{0\nu\beta\beta})^{-1}=G_{0\nu\beta\beta}(Q,Z)|\mathcal{M}_{0\nu\beta\beta}(A,Z)|^2|m_{\beta\beta}|^2~,
\label{eq:halflife}
\end{align}
where $G_{0\nu\beta\beta}(Q,Z)$ is the phase-space factor of the particular transition which depends on the isotope's $Q$ value and is well known, $|\mathcal{M}_{0\nu\beta\beta}(A,Z)|^2$ is the nuclear matrix element which currently presents a considerable source of theoretical uncertainty \cite{Engel:2016xgb,Agostini:2022zub,Belley:2023btr}, and $|m_{\beta\beta}|$ is the effective neutrino mass defined as \cite{Bilenky:1987ty}
\begin{align}
|m_{\beta\beta}|=\left|\sum_{i=1}^3 U_{ei}^2 m_i\right|\,.
\label{eq:mbb}
\end{align}
The effective neutrino mass contains the particle physics information of interest relevant for understanding neutrino masses and mixings and is the focus of this paper.
We write it in terms of mixing parameters in the standard parametrization of the neutrino mixing matrix \cite{ParticleDataGroup:2022pth}, the PMNS matrix \cite{Pontecorvo:1957cp,Maki:1962mu}, where we choose to assign the Dirac CP phase to the second row of the matrix such that eq.~\eqref{eq:mbb} is independent of it\footnote{It is clear from eq.~\eqref{eq:mbb} that there can be only two physical phases; the inclusion of $\delta$ in the first row leads to an additional phase redundancy which does not affect observables.}. In this case the PMNS matrix reads
\begin{widetext}
\begin{equation}
U_{\text{PMNS}}=\begin{pmatrix}
c_{12} c_{13} \text{e}^{\text{i}\alpha/2} &s_{12} c_{13} \text{e}^{\text{i}\beta/2} &s_{13} \\
(-c_{23} s_{12} - c_{12} s_{13} s_{23} \text{e}^{\text{i} \delta})\text{e}^{\text{i}\alpha/2}&(c_{12} c_{23} - s_{12} s_{13} s_{23} \text{e}^{\text{i}\delta})\text{e}^{\text{i}\beta/2} &c_{13} s_{23} \text{e}^{\text{i} \delta}\\
(-c_{12} c_{23} s_{13} +\text{e}^{-\text{i} \delta} s_{12} s_{23})\text{e}^{\text{i}\alpha/2}&(-c_{23}s_{12} s_{13} - \text{e}^{-\text{i} \delta} c_{12} s_{23})\text{e}^{\text{i}\beta/2}&c_{13}c_{23}
\end{pmatrix}\,,
\end{equation}
\end{widetext}
where we use the short hand notation $c_{ij}=\cos\theta_{ij}$, $s_{ij}=\sin\theta_{ij}$ and the Majorana phases $\alpha,\beta$ where one of them can be between $\in[0,\pi]$, the other one $\in[0,2\pi]$. Thus we can see that $|m_{\beta\beta}|$ is a function of seven free parameters: the three neutrino masses, two mixing angles and two Majorana phases \cite{Bilenky:1980cx,Schechter:1980gr,Doi:1980yb}.

A non-trivial feature of the seven parameters is that the predicted range of 0$\nu\beta\beta$ also depends on the neutrino mass ordering.
Oscillation experiments are starting to provide hints for the neutrino mass ordering, in particular, when combined in global fits which currently show a preference for the NO \cite{deSalas:2020pgw,Esteban:2020cvm,Capozzi:2021fjo}\footnote{The hints coming from long baseline accelerator neutrino experiments, however, might be an indication of new physics \cite{Kelly:2020fkv,Denton:2020uda,Chatterjee:2020kkm}.}.
Using the measured values of the neutrino mass splittings and mixing angles, $|m_{\beta\beta}|$ depends on only three unknown parameters: the absolute neutrino mass scale which is constrained to be at most somewhat light, and two Majorana phases which are completely unconstrained.
Note that $|m_{\beta\beta}|$ only constrains at most one combination of the two phases; unless $|m_{\beta\beta}|\approx0$ it is not possible to determine both Majorana phases using oscillation data with a detection of $0\nu\beta\beta$ alone \cite{Ge:2016tfx,Ge:2019ldu}. 
Furthermore, the Majorana phases do not lead to manifest CP violation in $0\nu\beta\beta$ \cite{deGouvea:2002gf,Barger:2002vy}\footnote{Note that the Majorana phases can lead to CP violating phenomena in other observables, for example in the leptogenesis scenario where a lepton asymmetry is generated via the decay of heavy, right-handed neutrinos which depends on the Majorana phases of these neutrinos \cite{Fukugita:1986hr}.}, they affect the $0\nu\beta\beta$ amplitude in a CP-even way, which excludes the possibility to determine them by considering $0\nu\beta\beta$ with the emission of two electrons and its CP conjugated process with the emission of two positrons.

\begin{figure*}
\centering
\includegraphics[width=0.49\textwidth]{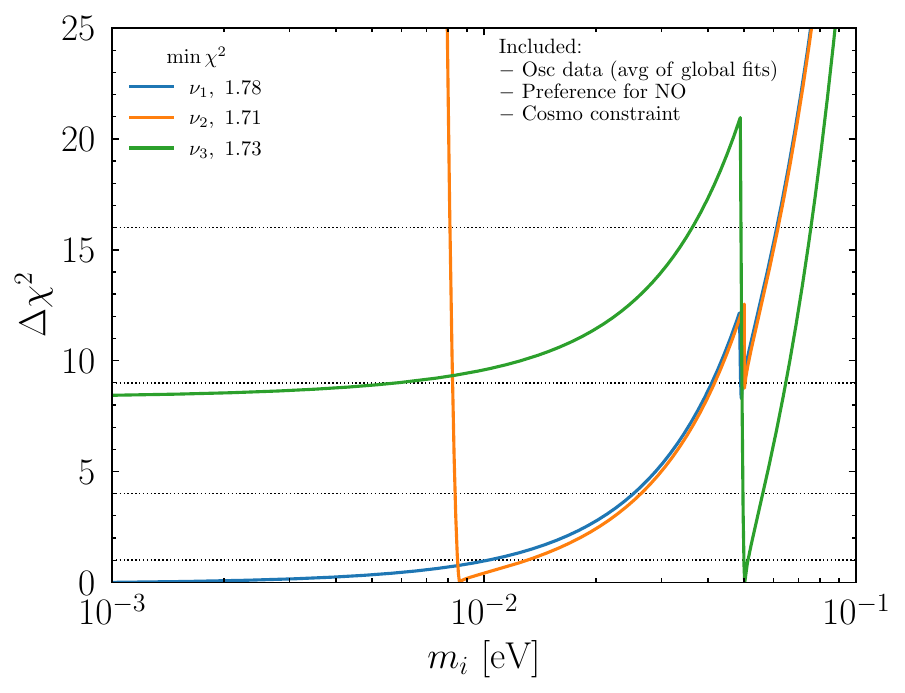}
\includegraphics[width=0.49\textwidth]{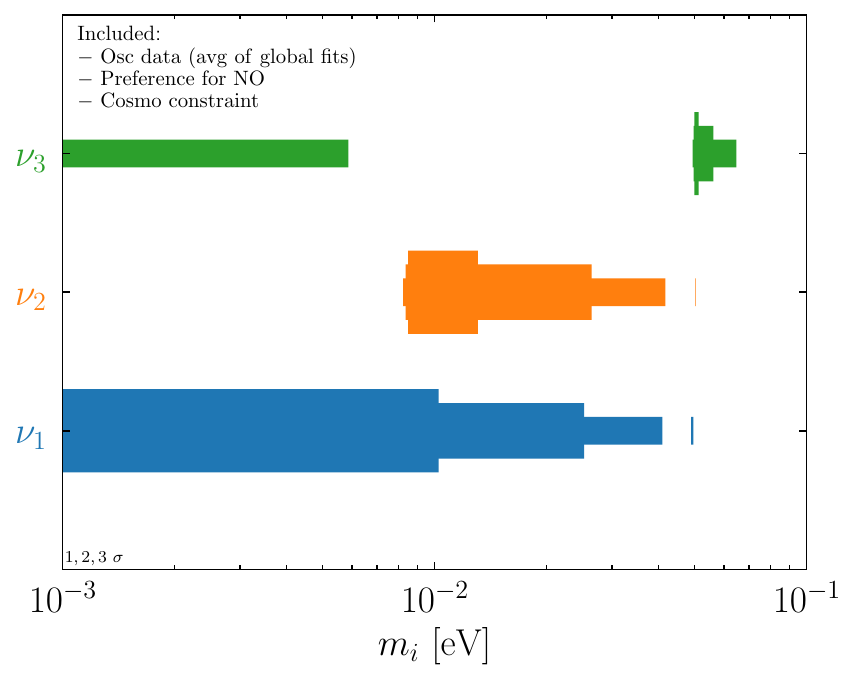}
\caption{The current knowledge on the absolute masses of the three neutrinos.
The data included is the average of the three global fits for $\Delta m^2_{21}$, $|\Delta m^2_{31}|$, and the preference for the normal ordering \cite{Esteban:2020cvm,deSalas:2020pgw,Capozzi:2021fjo}, as well as the cosmological constraint on the sum of the neutrino masses that does not yet show evidence for neutrino masses \cite{DiValentino:2021hoh}.
\textbf{Left}: The $\Delta\chi^2$ for parameter estimation, where we also note that the minimum $\chi^2$ for each state is $\sim1.75$ and thus is an acceptable fit to the data.
\textbf{Right}: The 1, 2, 3 $\sigma$ preferred regions for each mass state, individually.
The thicker regions are more preferred.}
\label{fig:absolute masses}
\end{figure*}

The absolute mass scale can, in principle, be constrained by beta decay end-point experiments such as KATRIN \cite{KATRIN:2019yun}, but the best constraints up to now come from cosmology.
Constraints vary from $\sum m_\nu<[87,90]$ meV at 95\% CL \cite{DiValentino:2021hoh,diValentino:2022njd}.
We take 90 meV \cite{DiValentino:2021hoh} as our fiducial number which then maps onto $m_1\lesssim17$ meV in the normal ordering for the best fit oscillation parameters. 
Current cosmological data seems to be incompatible with the inverted ordering at 95\% CL, although the details of this constraint depend considerably on one's choice of priors \cite{Gariazzo:2018pei,Gariazzo:2022ahe,Gariazzo:2023joe}.
The currently preferred regions for the neutrino masses are shown in fig.~\ref{fig:absolute masses} using oscillation data, the oscillation preference for the normal ordering (not used in the statistical tests elsewhere in this paper), and the cosmological constraint on the sum of neutrino masses (also not used in the statistical tests elsewhere in this paper).
Note that the different preferred regions for each mass are correlated with one another.
We see that $m_2$ has both an upper and lower limit while either $m_1$ or $m_3$ can  be zero.
We also see that each mass state has two disjoint preferred regions due to the different mass orderings as well as the important constraint from cosmology.

The current best limit on $|m_{\beta\beta}|$ is from KamLAND-Zen with $^{136}$Xe \cite{KamLAND-Zen:2022tow}
\begin{equation}
| m_{\beta\beta}|^{\text{exp}}<(36 – 156) \text{ meV}\,,
\end{equation}
where the range of values is due to the range of predictions for the nuclear matrix element. The most optimistic matrix element values indicate that this constraint starts to push into the inverted hierarchy while future experiments \cite{DAndrea:2021gqg} will further probe a large part of this region, subject to nuclear matrix element uncertainties.
In addition, future constraints on the neutrino mass scale from cosmology have important implications for $0\nu\beta\beta$ \cite{Ettengruber:2022mtm}.

\begin{figure}
\centering
\includegraphics[width=0.49\textwidth]{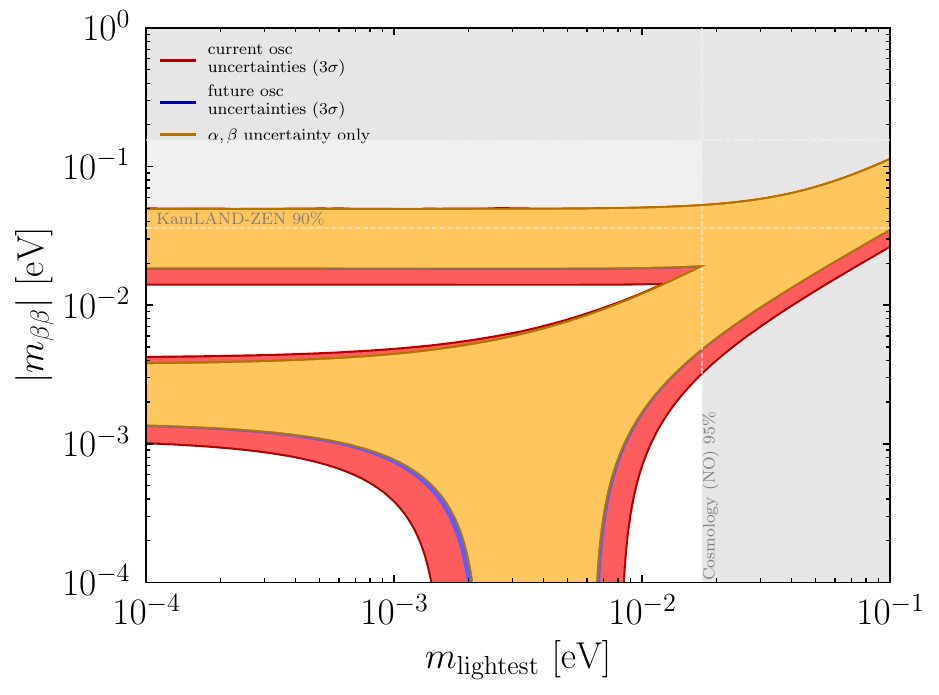}
\caption{The currently allowed region in the $|m_{\beta\beta}|-m_{\text{lightest}}$ plane for both mass hierarchies (the upper band corresponds to IO, the lower region to NO).
The yellow region is the expected allowed region with perfect knowledge of the oscillation parameters and the blue region indicates the increased region allowed due to the expected future precision in the oscillation parameters from DUNE and JUNO.
The red region indicates the increased region including the current uncertainties on the oscillation parameters.
All contours are drawn at true $3\sigma$.
The current upper limit on $|m_{\beta\beta}|$ from KamLAND-Zen is shown as horizontal gray bands where the darker and lighter gray regions assume different determinations of the nuclear matrix element \cite{KamLAND-Zen:2022tow}. The upper bound from cosmology on the absolute neutrino mass scale in NO \cite{DiValentino:2021hoh} is shown as a vertical gray band.}
\label{fig:mbb_allowed}
\end{figure}

In fig.~\ref{fig:mbb_allowed} we show the allowed regions in the $|m_{\beta\beta}|-m_{\text{lightest}}$ plane based on our knowledge of oscillation data, as well as the constraints on the lightest neutrino mass and upper limits on $|m_{\beta\beta}|$. The regions are drawn at the $3\sigma$ limit which means we impose that the total $\Delta\chi^2$, understood as the sum of all $\Delta\chi^2$ of the mixing angles and mass splittings, is equal to 11.83 which is $3\sigma$ with 2 dof\footnote{The choice of the number degrees of freedom is here is non-trivial.
Our choice is based on the fact that since there are two physics parameters: $|m_{\beta\beta}|$ and $m_{\rm lightests}$ this corresponds to two degrees of freedom.
While they are clearly related to each other, even with known oscillation parameters, the additional freedom from the two Majorana phases more than ensures that $|m_{\beta\beta}|$ is a distinct degree of freedom.}.
This is different from what is commonly done in the literature where each individual oscillation parameter is allowed to increase to some critical threshold without consideration for the total test statistic.
We do not impose information in the test statistic for the lightest mass from cosmological measurements or from $|m_{\beta\beta}|$, although since the latter only pushes into the inverted hierarchy it would not affect a discussion of the funnel.
We also do not include a penalty factor for current preference from the oscillation data for the normal ordering over the inverted ordering, although this also would not affect the funnel discussion.
We avoid those constraints because they make the distinction between the normal ordering and the inverted ordering complicated in a way that depends quite sensitively on the precise statistical test performed.
Finally, we do not include any information about $\delta$ (even though $\delta$ does not affect 0$\nu\beta\beta$, it is relevant for specific classes of flavor models) from long-baseline oscillation data as there is a mild tension among the two relevant data sets, NOvA and T2K \cite{Denton:2020uda,Chatterjee:2020kkm} and most values are allowed in any case.
The yellow regions show the allowed region if all oscillation parameters are known perfectly at the best fit values from \cite{Esteban:2020cvm}.
The only free parameters in the yellow region are the two Majorana phases.
The blue region shows the enlarged region that we can expect with the expected precision of the oscillation parameters from DUNE and JUNO \cite{JUNO:2022mxj,DUNE:2020ypp}.
This shows that the future measurements of the oscillation parameters by DUNE and JUNO will take us very close to the perfect knowledge case.
The red regions show the additional parameter space due to the current oscillation uncertainties, also taken from \cite{Esteban:2020cvm}.
We see that future oscillation experiments will constrain the parameter space further close to the relevant limit of perfect information, however the oscillation parameters are already measured rather precisely such that the Majorana phases present the largest uncertainty in the allowed regions of parameter space.
Therefore, predictions which are in agreement with the oscillation data but additionally predict the Majorana phases are of particular phenomenological interest as they prefer only parts of the generally allowed parameter space.

\begin{figure}
\centering
\includegraphics[width=1.8in]{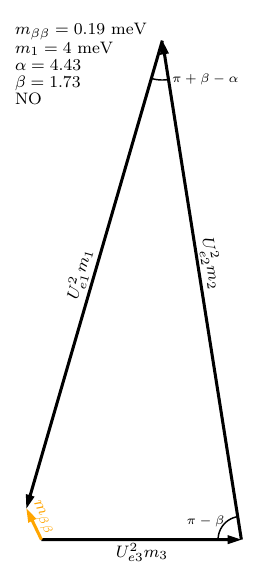}
\caption{A visual representation of $m_{\beta\beta}$ on the complex plane in the normal ordering for some choice of the Majorana phases and the mass of the lightest neutrino.
Since the three legs that make up $m_{\beta\beta}$ nearly close in this example, $|m_{\beta\beta}|$ is small enough to be in the funnel.}
\label{fig:triangle}
\end{figure}

We are specifically interested in the funnel region in NO, which we define as $|m_{\beta\beta}|<10^{-3}$ eV, consistent with other analyses in the literature, e.g.~\cite{Pascoli:2007qh,Ge:2016tfx,Ge:2019ldu}.
Such small values of $|m_{\beta\beta}|$ can only be achieved if the atmospheric mass ordering is normal and for $m_1\in[6\times 10^{-4},8\times 10^{-3}]$ eV assuming the best fit values of the neutrino parameters from oscillations \cite{Esteban:2020cvm}.
To understand the cancellation we interpret the expression for $|m_{\beta\beta}|$ as a quadrilateral in the complex plane, see fig.~\ref{fig:triangle} (see \cite{Vissani:1999tu} for an alternative graphical representation of $|m_{\beta\beta}|$).
If $|m_{\beta\beta}|\approx 0$, the quadrilateral reduces to a triangle.
Since $m_1|U_{e1}^2|$ grows faster with $m_1$ than $m_2|U_{e2}^2|$ and $m_3|U_{e3}^2|$ in the NO, there are values of $m_1$ where the sum or difference of $m_2|U_{e2}^2|$ and $m_3|U_{e3}^2|$ corresponds to $m_1|U_{e1}^2|$.
The situation is different in IO as the inequalities $m_1|U_{e1}^2|>m_2|U_{e2}^2|>m_3|U_{e3}^2|$ are satisfied for all values of $m_3$ and the currently allowed values for the mixing matrix elements from \cite{Esteban:2020cvm}. There are no values of the lightest mass where two sides of the quadrilateral sum up another side.
Therefore, the quadrilateral never collapses to a triangle and the minimum of $|m_{\beta\beta}|$ in the IO is $|m_{\beta\beta}|=19.8$ meV with $m_3=2.98$ meV. These values for the absolute neutrino mass scale which lead to $|m_{\beta\beta}|<10^{-3}$ eV can be tested with the next generation of laboratory based experiments like the ECHo experiment \cite{Gastaldo:2017edk}, Project 8 \cite{Project8:2017nal}, and the PTOLEMY experiment \cite{PTOLEMY:2019hkd} as well as cosmological experiments \cite{Font-Ribera:2013rwa,SimonsObservatory:2018koc,CMB-S4:2016ple,Brinckmann:2018owf} which will be sensitive down to neutrino masses in the $\mathcal{O}(10~\text{meV})$ region.
This, combined with the preference for the NO over the IO from oscillation data, makes a study of the funnel region most timely.

\section{Results for model category predictions}
\label{sec:results}
In this section we will introduce the categories of model predictions we study, provide an overview of the underlying theories, and present the preferred regions of parameter space and fractions in the funnel. 

We start by providing a complete phenomenological study of categories of models which make predictions for observables which enter $|m_{\beta\beta}|$.
These categories of models can be further subdivided into groups of model predictions which have the same predictions.
These groups of predictions are defined to cover existing individual models that are studied in the literature but are expanded to include other conceivable models with different combinations of the same input parameters. 
An important condition of our analysis is that we will not be concerned with whether or not all of these particular phenomenological predictions can be fully realized in concrete models, and we simply consider the option that they could be, thereby providing a phenomenological starting point for model builders by investigating conceivable model predictions.
 
\begin{figure}
\centering
\includegraphics[width=\columnwidth]{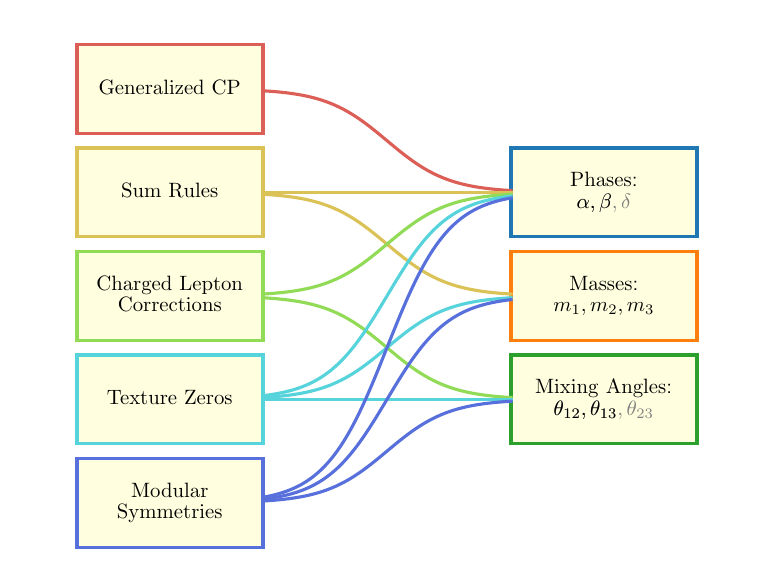}
\caption{Overview and categorization of the models studied and the parameters they predict.
The grayed parameters, $\delta$ and $\theta_{23}$, do not affect neutrinoless double beta decay, and not all flavor models (specifically sum rules) that predict the Majorana phases $\alpha$ and $\beta$ also predict $\delta$.
Each category contains groups of models for which we derive results.
The groups of models contain individual models realized in complete scenarios, for example based on underlying symmetries.}
\label{fig:model_overview}
\end{figure}

The model categories and the neutrino parameters they predict are schematically shown in fig.~\ref{fig:model_overview}.
Starting from the phenomenological point of view from model predictions for observables entering $|m_{\beta\beta}|$, we consider categories of models which make predictions for one or several of them.
Indeed, each phenomenological category makes predictions at a certain level in the mass matrix.
Predictions for the mixing parameters are generally driven by the structure of the neutrino mass matrix while predictions for the neutrino masses depend on the number of free parameters in the neutrino mass matrix. Models which affect the neutrino masses typically also predict a lower (and upper) limit on the lightest neutrino mass, allowing an additional probe of these categories via experiments sensitive to the absolute neutrino mass scale.

The five model categories we consider in this work are as follows:
\begin{itemize}
\item 
{\textbf{Generalized CP }}(\S~\ref{sec:gcp}) which makes predictions for all three complex phases only.
\item
{\textbf{Sum rules}} (\S~\ref{sec:sumrules}) which make predictions for the masses and the Majorana phases only\footnote{We will consider sum rules for the masses here. There exists another category of sum rules which involves the angles, these typically arise in models with charged lepton corrections, see sec.~\ref{sec:chargedlep}.}.
\item 
{\textbf{Charged lepton corrections}} (\S~\ref{sec:chargedlep}) which make predictions for the mixing angles and complex phases.
\item 
{\textbf{Texture zeros}} (\S~\ref{sec:tex0}) which make predictions for all nine parameters in the mass matrix.
\item {\textbf{Modular symmetries}} (\S~\ref{sec:modular}) which also make predictions for all nine parameters in the mass matrix.
\end{itemize}

To better understand the predictions of these models, we compare the number of constraints to the number of free parameters in the neutrino sector. As we are interested in neutrinoless double beta decay which can only happen for Majorana neutrinos, we focus on this case only, even though the model categories we study here also allow for Dirac neutrinos.
The complex, symmetric Majorana mass matrix has twelve parameters of which three phases can be absorbed into the three flavor eigenstates. Therefore, we are left with nine free parameters. The Majorana mass matrix $M_\nu$ is diagonalized as 
\begin{align}
M_\nu=U_{\text{PMNS}}~ D_\nu~ U_{\text{PMNS}}^T
\label{eq:diag}
\end{align}
with the PMNS matrix $U_{\text{PMNS}}$ \cite{Pontecorvo:1957cp,Maki:1962mu} and the diagonal mass matrix $D_\nu$ which contains the eigenvalues of $M_\nu$ which are the light neutrino masses including the Majorana phases $D_\nu=\text{diag}(m_1\text{e}^{\text{i}\alpha},~m_2\text{e}^{\text{i}\beta},m_3)$ amounting to nine free parameters.
In appendix \ref{sec:massmatrix_el}, we give the expressions for the mass matrix elements as a function of the mixing parameters and mass eigenvalues. Both sides of eq.~\eqref{eq:diag} are parametrized with the same number of parameters: nine.
Nevertheless, there is no one-to-one mapping of the mixing parameters to the matrix elements, all mass matrix elements depend on a combination of mixing parameters.

We point out that the nine parameters in the mass matrix need not be split up in terms of masses (eigenvalues), mixing angles, and complex phases in the usual way; there are other viable parameterizations of the degrees of freedom of the mass matrix.
One such example is with SU(3) generators (e.g.~Gell-Mann matrices), see appendix \ref{sec:gell-man}.

Focused on the usual parameterizations, the nine free parameters in the neutrino sector, assuming Majorana neutrinos, are the three neutrino masses, three neutrino mixing angles, and three CP phases. 
Out of these nine parameters, five have been measured at neutrino oscillation experiments (three angles, two mass splitting) while a sixth parameter, the Dirac CP phase will be measured in the future \cite{DUNE:2020ypp,Hyper-Kamiokande:2018ofw}.
Out of these five measured parameters, only four impact $0\nu\beta\beta$ as $|m_{\beta\beta}|$ does not depend on $\theta_{23}$ (it also does not depend on $\delta$).
Experiments sensitive to the absolute neutrino mass can constrain one parameter which also plays a role in $0\nu\beta\beta$.

Each different class of models not only impacts different sets of the physical parameters, as shown in fig.~\ref{fig:model_overview}, but also constrains those parameters at different levels.
Some, such as generalized CP or sum rules, provide only a small number of constraints while others, like modular symmetries, provide a large number of interconnected constraints among all the parameters.

\subsection{Numerical approach}
In order to quantify the validity of a given model and also its interplay with the funnel, we perform careful numerical studies, the methods of which are outlined here.
While there are some necessary choices to be made about the nature of the analyses, they have been made in such a way as to allow for a direct comparison among the different models and model classes and a representative numerical picture of the relationship between flavor models compatible with oscillation data and the funnel.

We study the predictions of the flavor models, requiring that the model predictions for the mixing parameters are in agreement with the experimental data; i.e.,~these flavor models correctly describe leptonic mixing and are hence not ruled out\footnote{We only consider priors on the three mixing angles and two mass splittings, but not on the sign of $\Delta m^2_{31}$, despite some evidence that it is positive. In case a model also predicts $\delta$ we do not include a prior. Similarly, we do not include any prior on the absolute neutrino mass scale.}. We will use the current global fit data for the mixing angles from \cite{Esteban:2020cvm} to derive the allowed values for $|m_{\beta\beta}|$.
As discussed in the previous section, we consider the true $3\sigma$ allowed regions of parameter space which corresponds to a total $\Delta \chi^2=11.83$ ($3\sigma$ for 2 dof) interpreted as the sum of the individual $\Delta\chi^2$ of the mixing angles and mass splittings. This approach is different from what is commonly done in the literature, where the allowed regions in the $|m_{\beta\beta}|-m_{\text{lightest}}$ plane (either in general or for a specific model) are derived by varying the individual mixing parameters in their $3\sigma$ ranges, which leads to a total $\Delta \chi^2$ larger than it should be.
This difference in the statistical approach leads to a difference in our results compared to results in the literature with other allowed regions being artificially looser than the quoted statistical significance implies.
Another crucial difference arises from our usage of up-to-date global fit results of the mixing parameters. In particular, the uncertainty on $\Delta m_{31}^2$ decreased in the past 5 years from 4\% to 1\%; after 2013 the uncertainty on $\Delta m^2_{21}$ and $\theta_{12}$ remained similar, and the improvements in the precision of $\theta_{13}$ \cite{Denton:2022een} do not have a huge impact on the uncertainties for 0$\nu\beta\beta$.

To determine the fraction of models in the funnel we follow several steps for each class of models:
\begin{enumerate}
\item We first calculate the number of models which are viable. These are the models that are in agreement with the oscillation data.
\item
Then we determine which of those have any fraction within the funnel which we define to be $m_{\beta\beta}<10^{-3}$ eV.
\item
Then we determine the fraction of each model that is within the funnel as outlined below.
\end{enumerate}

Different classes of models structure their predictions differently; some provide constraint equations while others also introduce new underlying parameters of the model.
Thus, there is not a straightforward means to consistently sample the model space; a study of one model (or one class of models) might prefer a different statistical test and come to slightly different conclusions.
Instead, we use a simple phenomenologically motivated definition that will be equally representative for all models, although we caution the reader that, even still, some regions of parameter space may be over-/under-represented compared to the representative size of the underlying parameters.
We define the fraction within the funnel as,
\begin{equation}
f=\frac{\int_{\rm funnel} d\log m_{\rm lightest}d\log m_{\beta\beta}}{\int d\log m_{\rm lightest}d\log m_{\beta\beta}}\,,
\label{eq:funnel fraction}
\end{equation}
where the integrals are over the allowed parameter space for a given model. For the denominator we only take the NO into account because the mass ordering will be known at high significance before neutrinoless double beta decay experiments probe the normal ordering.
We also consider the same expression with a linear distribution on the masses ($d\log m\to dm$).
In addition, we bound the integral $m_{\rm lightest}\in[10^{-4},10^{-1}]$ eV and $m_{\beta\beta}\in[10^{-4},10^0]$ eV as shown in fig.~\ref{fig:mbb_allowed}.
In some cases this affects the numerical results somewhat artificially; however, these numbers are well motivated by existing limits on the lightest neutrino mass and $|m_{\beta\beta}|$ and the general narrative does not change much.
In addition to the fraction within in the funnel we also show probability density functions (PDFs) of each category in the $|m_{\beta\beta}|-m_\text{lightest}$ plane where the darker the color, the higher the PDF. Due to the common choice to present these plots in log-log scale the regions covered in these models are not necessarily uniform in the colored regions.

We now turn to the five model classes in the following subsections.

\subsection{Models with generalized CP} 
\label{sec:gcp}
\begin{figure*}
\centering
\includegraphics[width=0.49\textwidth]{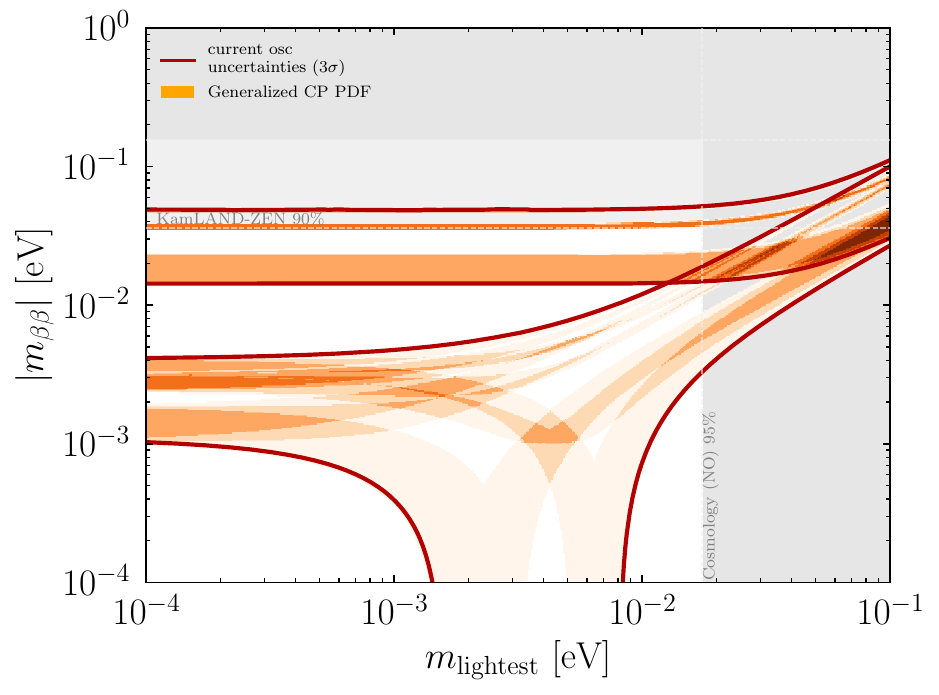}
\includegraphics[width=0.49\textwidth]{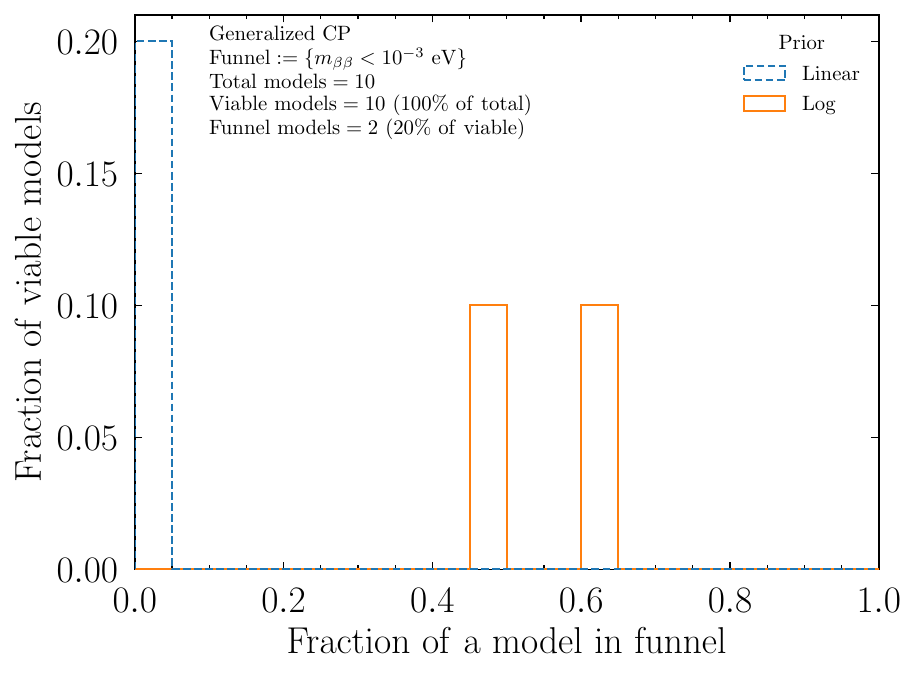}
\caption{\textbf{Left}: The PDF of all 10 different models within the generalized CP classification using the latest constraints from the oscillation data.
\textbf{Right}: The histograms showing how much of each model is in the funnel.
We see that all 10 models are consistent with oscillation data and two of them are in the funnel.
Using a log prior on $m_{\rm lightest}$ there is a $\sim50\%$ probability that they are in the funnel, while with a linear prior the probability is much less.}
\label{fig:gCP results}
\end{figure*}

In models where CP is a conserved quantity the values of the 
CP violating phases are constrained to be 0 or $\pi$ \cite{Wolfenstein:1981rk,Kayser:1984ge,Bilenky:1984fg,Branco:1986gr,Feruglio:2012cw,Holthausen:2012dk}. On the other hand, the phases can have non-trivial phases if a discrete symmetry is combined with a generalized
CP symmetry \cite{Feruglio:2012cw}. Apart from the CP conserving values, possible predictions for the Majorana phases are 
$\pi/2,~3\pi/2$ \cite{Ding:2013nsa,Ding:2014hva,King:2014rwa,Ding:2014ssa,Hagedorn:2014wha,Ding:2014ora,Ding:2015rwa}. 
Similar to \cite{Penedo:2018kpc} we consider 16 combinations of values for the Majorana phases $(\alpha,\beta)\in\{0,\pi/2,\pi,3\pi/2\}$.
Out of the 16 combinations, several map onto each other (see appendix \ref{sec:independentgCP}) such that there are only 10 independent combinations.
All of them are viable since they only predict the Majorana phases and the ones with $(0,~\pi),~(\pi,~0)$ predict a region in the funnel, see fig.~\ref{fig:gCP results}.
We find a $\sim50\%$ probability (using a log prior) that these two models are in the funnel.
Furthermore, these models cover much of the whole allowed region for $m_{\beta\beta}$.
Since the PDF is not uniform, however, there is a preference in these models for $m_{\beta\beta}$ values close to the lower allowed bound in IO and towards small values of the lightest mass in NO, even though all models are compatible with all values of $m_\text{lightest}$ as they do not predict a lower limit on the absolute neutrino mass.

\subsection{Models with mass sum rules}
\label{sec:sumrules}
Mass sum rules are relations between the three complex neutrino eigenvalues $m_i\text{e}^{\text{i}\alpha_i}$ (for overviews see \cite{Barry:2010yk,King:2013psa,Gehrlein:2020jnr}), and they arise in flavor models where the neutrino mass matrix depends on two complex parameters only \cite{Gehrlein:2017ryu}. Then the
three eigenvalues of the mass matrix are not independent but are related by a sum rule.
As one complex neutrino mass eigenvalue can be expressed as a function of the other two, these models constrain two parameters in the mass matrix and therefore these models predict two parameters of interest in $|m_{\beta\beta}|$.

Mass sum rules 
can be parameterized with 5 free real model parameters
\begin{align}
 c_1 \text{e}^{\text{i} \chi_1} ( m_1\text{e}^{\text{i}\alpha})^d+c_2 \text{e}^{\text{i} \chi_2} ( m_2\text{e}^{\text{i}\beta})^d+m_3^d=0\,,
 \label{eq:msr}
\end{align}
where $d$ is the power of the sum rule, $c_1, c_2$ are the real coefficients of the sum rule, and $\chi_1, \chi_2$ are the phases.
Note that we have set the coefficient and phase of $m_3$ to be 1 and 0, respectively.
 
Up to now 12 mass sum rules have been identified in over 60 different models \cite{Barry:2010yk,Bazzocchi:2009da,Ding:2010pc,Ma:2005sha,Ma:2006wm,Kang:2015xfa,Honda:2008rs,Brahmachari:2008fn,Altarelli:2005yx,Chen:2009um,Chen:2009gy,Cooper:2012bd,Altarelli:2009kr,Altarelli:2008bg,Hirsch:2008rp,Bazzocchi:2009pv,Everett:2008et,Boucenna:2012qb,Mohapatra:2012tb,Altarelli:2005yp,Altarelli:2006kg,Ma:2006vq,Bazzocchi:2007na,Bazzocchi:2007au,Lin:2008aj,Ma:2009wi,Ciafaloni:2009qs,Bazzocchi:2008ej,Feruglio:2013hia,Chen:2007afa,Ding:2008rj,Chen:2009gf,Feruglio:2007uu,Merlo:2011hw,Luhn:2012bc,Fukuyama:2010mz,Ding:2013eca,Lindner:2010wr,Hashimoto:2011tn,Ding:2011cm,Morisi:2007ft,Adhikary:2008au,Lin:2009bw,Csaki:2008qq,Hagedorn:2009jy,Burrows:2009pi,Ding:2009gh,Mitra:2009jj,delAguila:2010vg,Burrows:2010wz,Ahn:2014zja,Karmakar:2014dva,Ahn:2014gva,He:2006dk,Berger:2009tt,Kadosh:2010rm,Lavoura:2012cv,King:2012in,Adulpravitchai:2009gi,Dorame:2011eb,Dorame:2012zv}.
These previously studied sum rules have parameters within certain typical ranges: $c_1,~c_2\sim \mathcal{O}(1),~d=\pm1,\pm 1/2, \chi_{1},\chi_2=0,\pi,\pm \pi/2$. However other values for these parameters are possible. 

For this study we will remain agnostic about the model realizations of the mass sum rules and study mass sum rules
with $c_1,~c_2\in[1/6,2/6,3/6,4/6,5/6,1,2]$, $d=\pm 1,~\pm 1/2$, $\chi_{1},\chi_2=0,\pi, \pi/2, 3\pi/2$. These choices include 11 realized mass sum rules; additionally, we consider $c_1=2/(\sqrt{3}+1),~c_2=(\sqrt{3}-1)/(\sqrt{3}+1),~d=1,~\chi_1=0,~ \chi_2=\pi$ to fully cover the parameter space of allowed mass sum rules with constant coefficients.
For the sake of concreteness we do not include $d=\pm 1/4, \pm 1/3$  as they have not appeared yet in realized sum rules in the literature.
Our choice of parameters to study covers existing models in the literature.
It is conceivable that other models could also be realized with ratios of larger integers; in order to retain some amount of predictivity, we truncate the parameters at the level of existing models in the literature.

\begin{figure}
\centering
\includegraphics[width=0.49\textwidth]{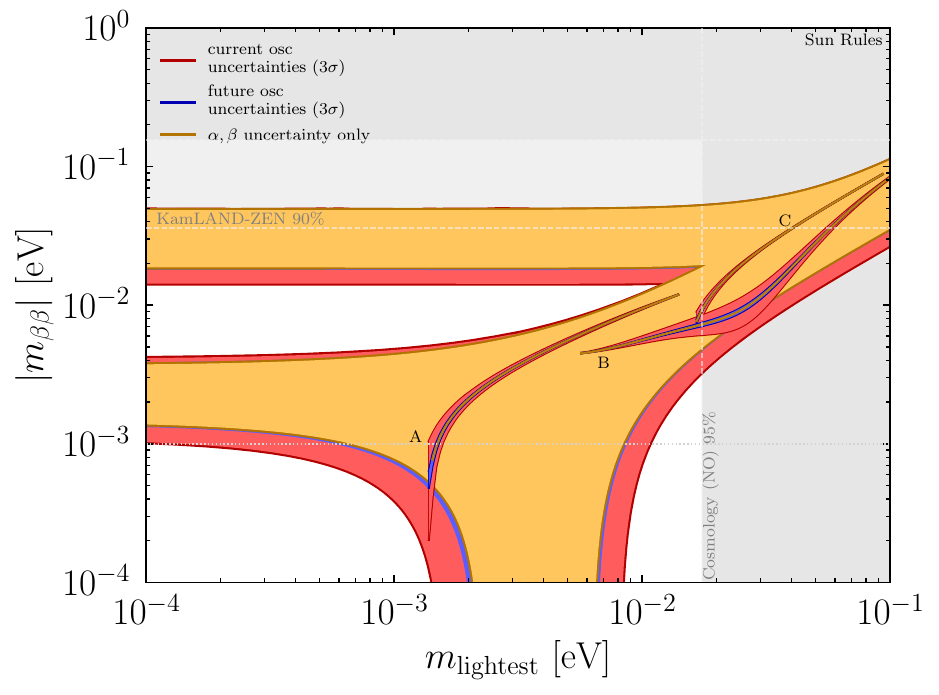}
\caption{The allowed region at $\Delta\chi^2=11.83$ for the current constraint on the oscillation parameters is shown in red, the region for the expected future precision at $\Delta\chi^2=11.83$ is shown in blue, and perfect precision is shown in orange.
The large regions are with no model constraints while the smaller regions are for various sum rules $(c_1,c_2,d,\chi_1,\chi_2)$.
A: $(1,2,\frac12,\pi,\frac\pi2)$, B: $(\frac12,\frac12,-\frac12,\pi,\pi)$, and C: $(1,2,1,\pi,0)$.}
\label{fig:several}
\end{figure}

A mass sum rule can be interpreted as a triangle in the complex plane which closes if the sum rule is fulfilled, this leads to prediction for the Majorana phases depending on the light neutrino masses.
Furthermore, there is a lightest neutrino mass for which the triangle can close, in some cases there is also an upper limit on the neutrino masses.
For some coefficients the mass sum rule can never be fulfilled like $-2\sqrt{m_1 \text{e}^{\text{i}\alpha}}+1/2\sqrt{m_2\text{e}^{\text{i}\beta}}-\sqrt{m_3}=0$, while in other cases the mass sum rule can only be fulfilled for one neutrino mass ordering but not for the others, like $m_1 \text{e}^{\text{i}\alpha}-2 m_2\text{e}^{\text{i}\beta}-m_3=0$ which can only be fulfilled in the NO.
All of these predictions affect $|m_{\beta\beta}|$, making this observable the ideal probe of the existence and type of mass sum rules. In general, mass sum rules only allow a small range in the $|m_{\beta\beta}|-m_{\text{lightest}}$ parameter space \cite{Cirigliano:2022oqy}.
In fig.~\ref{fig:several} we show the allowed ranges for several representative sum rules. We see that predictions from sum rules can be very different, and while all sum rules predict a lower bound on the lightest mass, some of them also predict an upper bound.

\begin{figure*}
\centering
\includegraphics[width=0.49\textwidth]{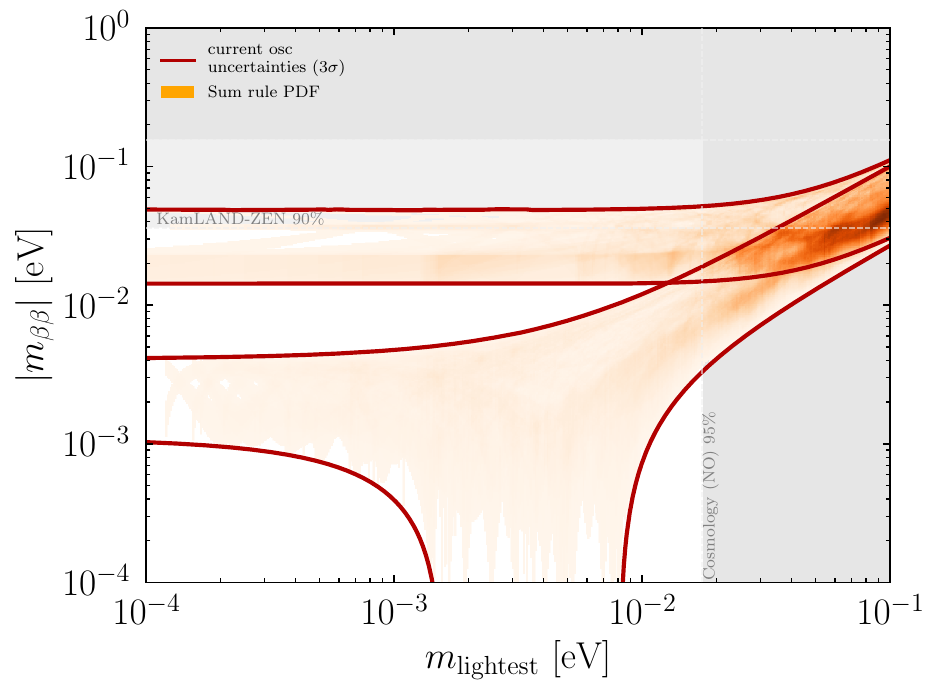}
\includegraphics[width=0.49\textwidth]{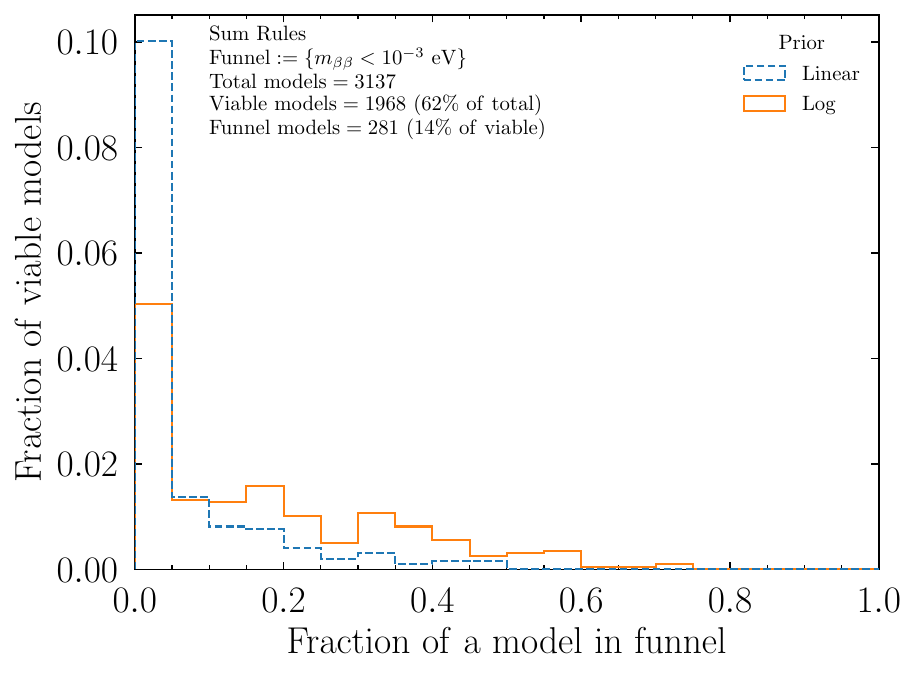}
\caption{The same as fig.~\ref{fig:gCP results} but for sum rules.}
\label{fig:SR results}
\end{figure*}

In fig.~\ref{fig:SR results} we show the PDF of models with sum rules.
Out of the 3137 models, 1968 are viable of which 14\% are in the funnel. None of the 12 models realized in the literature has a fraction in the funnel\footnote{This statement seemingly contradicts previous results \cite{King:2013psa,Gehrlein:2016wlc} however this discrepancy arises due to the different choice of $\chi^2$ contours.}.
There are 17 models with at least 50\% in the funnel and they are enumerated in appendix \ref{sec:msrfunnel}\footnote{We provide a text file containing all sum rule models at \href{https://peterdenton.github.io/Data/0nubb\_Survey/index.html}{peterdenton.github.io/Data/0nubb\_Survey}}.

From fig.~\ref{fig:SR results} we also see that sum rules cover the whole parameter space rather uniformly however none of the models we studied allows for $m_1<10^{-4}$ eV in NO while there is no lower bound in IO. This can be understood as in NO there is a hierarchy between the masses, even for small $m_1$ which requires larger coefficients than we study to fulfill the sum rule for smaller masses.
In IO, on the other hand, $m_1$ and $m_2$ are nearly degenerate such that cancellations between them can occur, which allows the sum rule to be fulfilled.

\begin{figure*}
\centering
\includegraphics[width=0.49\textwidth]{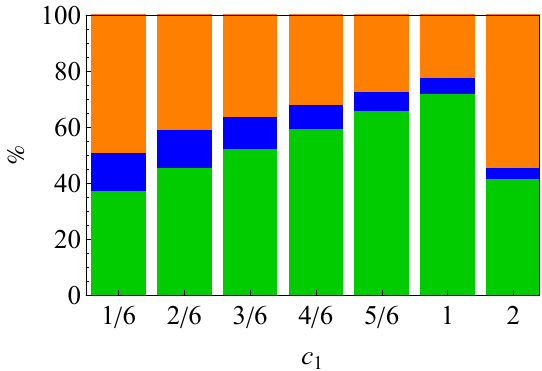}
\includegraphics[width=0.49\textwidth]{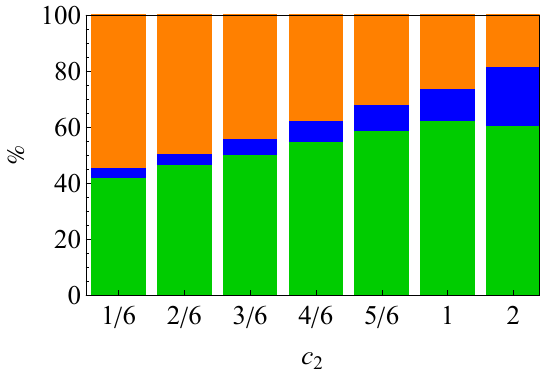}
\includegraphics[width=0.49\textwidth]{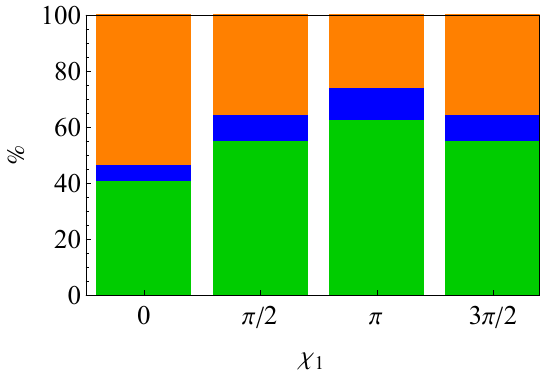}
\includegraphics[width=0.49\textwidth]{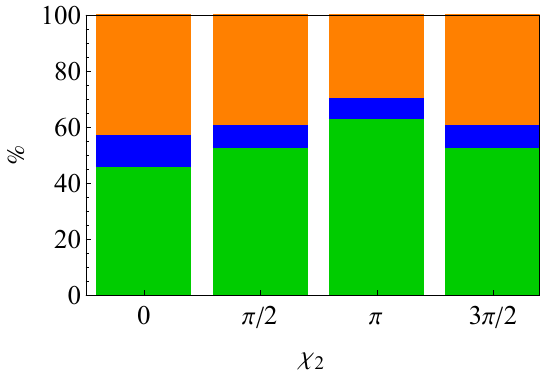}
\includegraphics[width=0.85\textwidth]{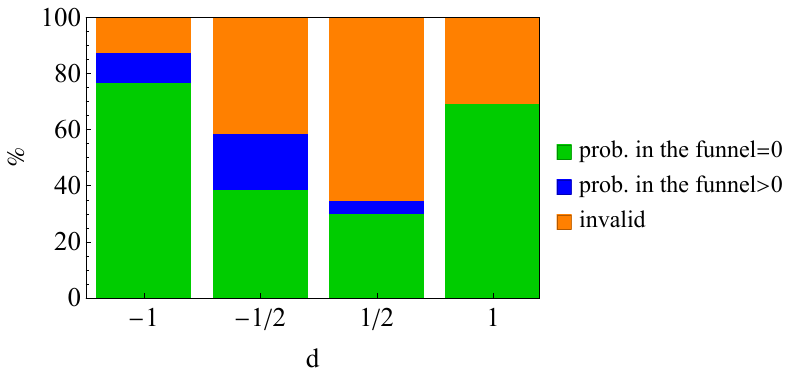}
\caption{The 3137 considered sum rule models split into the values of each of the five parameters with one panel for each of the five parameters.
The bars indicate the fraction of models with a specific value of one parameter that are either inconsistent with oscillation data (orange), consistent with oscillation data but never in the funnel (green), or consistent with oscillation data and some fraction in the funnel (blue). We only study the case of NO.
}
\label{fig:SR parameters}
\end{figure*}

We further investigate the roles the five individual sum rule parameters play in the behavior of the models as shown in fig.~\ref{fig:SR parameters}. 
These figures show, for each value of each parameter, the fraction of all models that are either not consistent with oscillation data (orange), consistent with oscillation data but never in the funnel (green), or consistent with oscillation data and some fraction in the funnel (blue).
Interesting trends appear.
We see that sum rule models are more likely to be in the funnel for small $c_1$ and large $c_2$.
The exponent $d$ also plays an important role, particularly that $d=1$ is never in the funnel as in this case the sum rule always leads to values of $m_3$ so large that the quadrilateral for $m_{\beta\beta}$ cannot collapse to a triangle. On the other hand, models with $d=-1/2$ have the largest fraction ($\approx$ 20\%) in the funnel. The values of $\chi_1,~\chi_2$ individually do not drastically impact the validity or fraction in the funnel, while we find that for $d=-1$ over 80\% of the models are viable but less than 40\% for $d=1/2$. 

We generally find that more than 50\% of the models studied are viable.
For the coefficients $c_1,~c_2$ we find that the larger they are, the more viable models we find; however, $c_1=2$ again leads to fewer viable models.

\begin{figure*}
\centering
\includegraphics[height=2.45in]{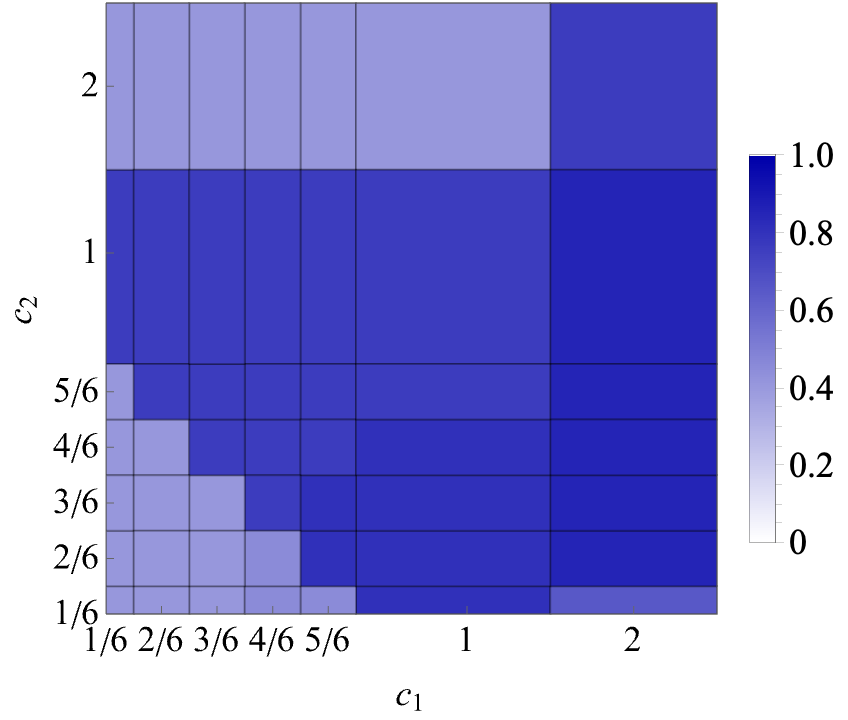}
\includegraphics[height=2.45in]{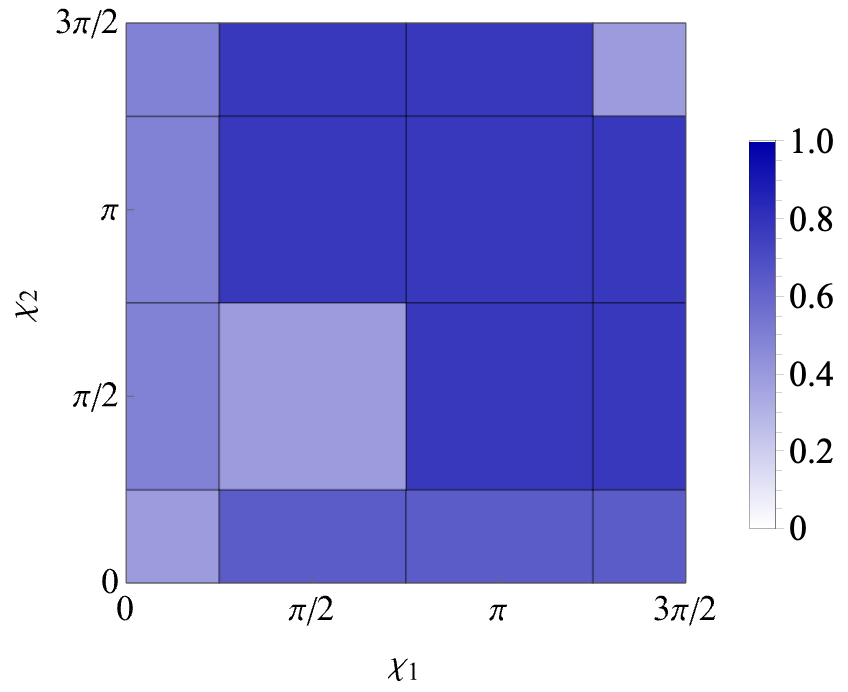}
\caption{Fraction of valid models with sum rules for a set of two model parameters where we find correlations.
}
\label{fig:SR2d}
\end{figure*}

Furthermore, if both coefficients are small, there is only a small fraction of valid models, which we also show in fig.~\ref{fig:SR2d}. Finally, even though the values of $\chi_1,~\chi_2$ individually are not very important for the validity or fraction in the funnel, we find a correlation between them, and larger values of both are preferred to find valid models, see fig.~\ref{fig:SR2d}.
For the other parameters we do not find strong correlations among them.

Thus, if the data indicates that we could be in the funnel or if one wants to build specific models that map onto sum rules that are consistent with current oscillation data and are or are not in the funnel, this can give some guidance about what kinds of parameters are likely to achieve those goals.

\subsection{Models with discrete symmetries in the neutrino sector and non-zero charged lepton mixing}
\label{sec:chargedlep}
Many flavor models based on discrete symmetries predict $\theta_{13}=0$ which is in strong contrast to the experimental data which prefers $\theta_{13}^{\text{exp}}\approx 8.5^\circ$ \cite{DayaBay:2018yms,RENO:2015ksa}. Therefore these predictions from discrete symmetries need to be corrected. A way to do so is by introducing a non-diagonal charged lepton mixing matrix as the measureable PMNS matrix is the product of the neutrino mixing matrix and the charged lepton mixing matrix $U_{\text{PMNS}}=U_e^\dagger U_\nu$. 
The introduction of a non-diagonal charged lepton mixing matrix leads to relations between the observable mixing parameters, including the Majorana phases. These relations are called mixing sum rules \cite{Ge:2011ih,Ge:2011qn,Marzocca:2013cr,Petcov:2014laa, Girardi:2014faa, Girardi:2015zva, Girardi:2015vha,Girardi:2016zwz,Girardi:2015rwa,Gehrlein:2022nss} (for reviews, see \cite{King:2013eh,King:2014nza,Petcov:2017ggy}) and are similar to the relations between the mixing parameters which arise in models with modular symmetries described below. A non-diagonal charged lepton mixing matrix could, for example, originate in grand unified theories based on SU(5) \cite{Georgi:1974sy} or SO(10) \cite{Georgi:1974my,Fritzsch:1974nn} where the structures of the mass matrices for the charged lepton mass matrix and down quarks coincide \cite{Antusch:2011qg,Marzocca:2011dh,Antusch:2009gu,Antusch:2012fb} such that the charged lepton sector exhibits CKM-like mixings \cite{Datta:2005ci}.
In \cite{Petcov:2014laa,Girardi:2016zwz} a detailed, systematic study of various forms of $U_\nu,~U_e$ in flavor models has been conducted, and the expressions for the Majorana phases, as well as for the mixing parameters, have been derived.

We will consider the cases of two or three rotations in the neutrino sector and one or two rotations in the charged lepton sector. For the neutrino mixing angles, we use $\theta_{23}^\nu=45^\circ$, and several cases for $\theta_{12}^\nu$ motivated by different popular symmetry forms of the neutrino mixing matrix, i.e.~$\sin^2\theta_{12}^\nu=1/3$, $1/2$, $1/(2+\phi_g)$, $(3-\phi_g)/4$, and $1/4$ 
with $\phi_g=(1+\sqrt{5})/2$ the golden ratio. We call these models TBM, BM, GRA, GRB, and HG standing for tri-bi-maximal mixing, bi-maximal mixing, golden ratio A form, golden ratio B form, and hexagonal form respectively.
Additionally, we consider models with three neutrino rotations with $\theta_{13}^\nu=\pi/10$, $\pi/20$, and $\arcsin(1/3)$, which we call T13-1, T13-2, and T13-3 respectively, motivated by existing models in the literature \cite{Bazzocchi:2011ax,Rodejohann:2014xoa,deAdelhartToorop:2011nfg,Ding:2012xx,King:2012in}.
The rotations in the charged lepton sector are free and are effectively constrained by the measured mixing angles.
In fact, for models with $\theta_{13}^\nu=0$ the charged lepton corrections are crucial to reproduce the observed mixing angles. However, charged lepton corrections also impact the predictions for the other mixing angles such that deviations from maximal $\theta_{23}$ can also be achieved. Therefore we also include models with two charged lepton rotations. However, we constrain ourselves to a maximum of a total of four rotations split between the neutrino and charged lepton sector as they provide sufficient freedom to reproduce the three measured mixing angles. More rotations or different predicted values of the 
neutrino or charged lepton mixing angles might arise, however, in concrete models \cite{Frampton:2004ud,deMedeirosVarzielas:2017sdv,Bernigaud:2022sgk}. 

To summarize the previous paragraph, we consider one or two charged lepton corrections where we study all combinations of $\theta_{13}^e,~\theta_{12}^e,~\theta_{23}^e$ rotations\footnote{We do the rotations in the standard order $U_\text{PMNS}=(U_{12}^e)^\dagger (U_{13}^e)^\dagger(U_{23}^e)^\dagger\Psi U_{23}^\nu U_{13}^\nu U_{12}^\nu Q$.}. Additionally, we also consider phases in the charged lepton mixing matrix which are necessary to obtain predictions on the phases in the PMNS matrix. We introduce the phase matrices $\Psi=\text{diag}(1,~\text{e}^{-\text{i}\psi_1},~\text{e}^{-\text{i}\psi_2})$ and $Q=\text{diag}(1,~\text{e}^{\text{i}\xi_1/2},~\text{e}^{\text{i}\xi_2/2})$. 
Explicitly, we derive results for the following scenarios
\begin{itemize}
\item two rotations in the neutrino sector, one charged lepton rotation (15 cases)
\begin{align}
U_{\text{PMNS}}&=(U_{ij}^e)^\dagger \Psi U_{23}^\nu(\pi/4) U_{12}^{\nu} ( \theta_{12}^{\nu,k})Q\nonumber\\&\text{where } (ij)\in\{12,~13,~23\}\nonumber\\
&\text{and }k\in \text{\{TBM, BM, GRA, GRB, HG\}}
\end{align}

\item two rotations in the neutrino sector, two charged lepton rotations (15 cases)
\begin{align}
U_{\text{PMNS}}&=(U_{ij}^e)^\dagger (U_{lm}^e)^\dagger\Psi U_{23}^\nu(\pi/4) U_{12}^{\nu} ( \theta_{12}^{\nu,k})Q\nonumber\\&\text{where } (ij)\in\{12,~13\},~ (lm)\in\{13,~23\},~(ij)\neq(lm) ~\nonumber\\
&\text{and }k\in \text{\{TBM, BM, GRA, GRB, HG\}}
\end{align}

\item three rotations in the neutrino sector, one charged lepton rotation (45 cases)
\begin{align}
U_{\text{PMNS}}&=(U_{ij}^e)^\dagger \Psi U_{23}^\nu(\pi/4) U_{13}^\nu(\theta_{13}^{\nu,p} ) U_{12}^{\nu} ( \theta_{12}^{\nu,k})Q~\nonumber\\&\text{where } (ij)\in\{12,~13~,~23\}
\\&\text{and }k\in \text{\{TBM, BM, GRA, GRB, HG\}}\nonumber
\\&\text{and }p\in \text{\{T13-1, T13-2, T13-3\}}\nonumber
\end{align}
\end{itemize}

For the phases contained in $\Psi,~Q$ we remain agnostic about their values and we vary them freely\footnote{Simultaneously employing, for example, a generalized CP symmetry allows to fix the values of the phases like in \cite{Girardi:2016zwz}.}.
Notice that all four phases are not physical in all cases.
In fact, only for the case of two charged lepton rotations with $\theta_{12}^e,\theta_{13}^e\neq0$ do all four phases play a role. The number of free parameters thus varies in the different scenarios.
The case of three neutrino rotations and one charged lepton rotation has the same number of rotations as the case with two of each; however, since the free mixing angles are always contained in the charged lepton sector, the case of two neutrino rotations and two charged lepton rotations has the most freedom.

These models make predictions for the mixing angles and the Majorana phases; therefore, they can be tested in 0$\nu\beta\beta$ experiments. 
In addition, these models also predict the CP phase $\delta$; however, we do not include a prior on $\delta$ in our analysis. Nevertheless, this prediction also presents a crucial test of this class of models \cite{Petcov:2014laa, Girardi:2014faa}.
On the other hand, these models do not predict a lower bound on the lightest neutrino mass.

For the models with two rotations in the neutrino sector and one charged lepton rotation, we find that 8 out of 15 models are viable. The BM mixing pattern in the neutrino sector cannot be brought into agreement with experimental data with one charged lepton rotation.
Other mixing patterns are viable assuming a 1-2 or 1-3 rotation in the charged lepton sector.
All models studied with two charged lepton rotations are viable.
In particular, for BM mixing, two charged lepton rotations are required to correct both vanishing $\theta_{13}$ and maximal $\theta_{12}$.
For three neutrino rotations and one charged lepton rotation, we find that only 8 out of 45 models are viable.

These results are in general agreement with results from the literature \cite{Girardi:2015vha,Girardi:2016zwz}. 
Nevertheless, we notice that improved precision on the oscillation parameters in comparison to the time where these studies were done now disfavors some models which were previously allowed.
In total we find that out of the 75 cases, 31 models are viable.

\begin{figure*}
\centering
\includegraphics[width=0.49\textwidth]{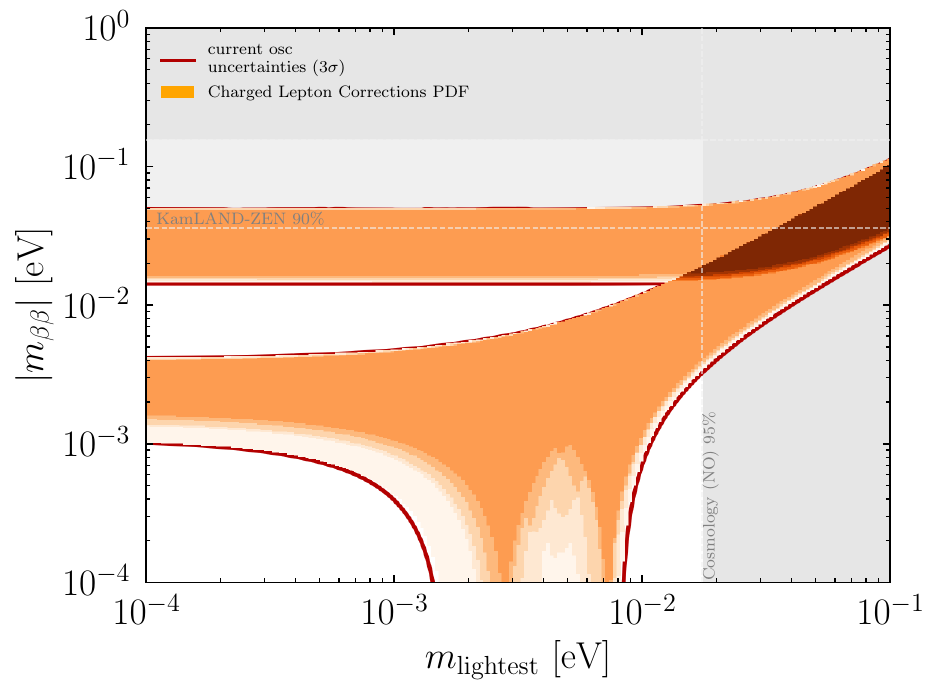}
\includegraphics[width=0.49\textwidth]{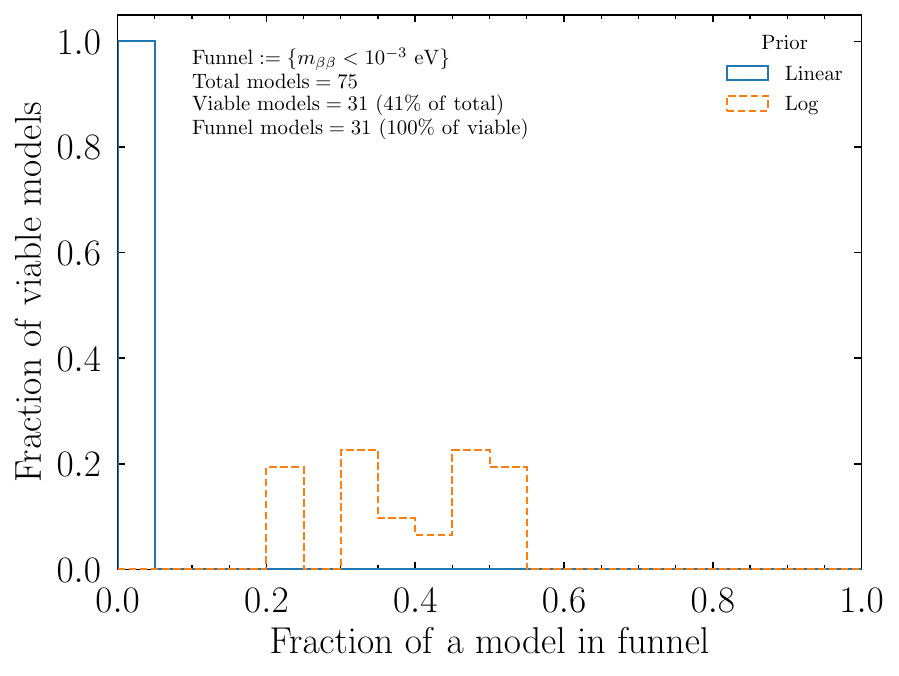}
\caption{The same as fig.~\ref{fig:gCP results} but for models with discrete symmetries and charged lepton corrections.}
\label{fig:CLC results}
\end{figure*}

The predictions for $0\nu\beta\beta$ experiments are shown in fig.~\ref{fig:CLC results}.
We find that many models predict a region in the funnel.
As this category of models does not predict the mass scale, the regions extend to small masses and cover the quasi-degenerate region disfavored by cosmology as well.
The funnel fractions are very similar in all models and between 20\% and 50\%, demonstrating that, in this category of models, up to a third of the parameter space can be contained in the funnel\footnote{We provide a text file containing all the charged lepton correction models at \href{https://peterdenton.github.io/Data/0nubb\_Survey/index.html}{peterdenton.github.io/Data/0nubb\_Survey}}. 

\subsection{Models with texture zeros}
\label{sec:tex0}
In models with texture zeros it is assumed that the complex symmetric Majorana mass matrix has some vanishing entries\footnote{Note that one can also consider the case where there are zeros in the charged- and neutral-lepton mass matrices separately \cite{Ramond:1993kv,Benavides:2022hca}; we will not consider these scenarios.}.

Of particular importance for this paper is the 1-1 element of the Majorana mass matrix which coincides with the observable $|m_{\beta\beta}|$, see
\cite{Merle:2006du,Lashin:2011dn, Dev:2006qe, Verma:2020gpl,Xing:2003jf,Xing:2003ic,BenTov:2011tj,Liu:2012axa}.
A symmetry realization of texture zeros can come from an extended scalar sector and suitable Abelian symmetries \cite{Grimus:2004hf}.
Here, however, we will remain agnostic of any underlying symmetry behind texture zeros as well as about the origin of the neutrino mass term\footnote{There are other models which constrain the number of free parameters in the mass matrix by imposing that the trace or the minor of the mass matrix is zero \cite{Branco:2002ie,Black:2000bk,Singh:2018tqu,Dey:2023tsk} which we won't consider here.}.

Majorana mass matrices with three or more independent texture zeros are already ruled out by current oscillation data \cite{Frampton:2002yf}, as in this case there are more observables than free parameters. Therefore, we will focus on one- and two-texture zero mass matrices\footnote{The case with no texture zeros is not predictive as there are more free parameters than observables.}.

For the vanishing mass matrix element $M_{\alpha\beta}=0$, the condition
\begin{align}
\sum_{i=1}^3 U_{\alpha i}U_{\beta i}D_i=0\,,
\label{eq:onezero}
\end{align}
applies, where $D_i$ stands for the elements of the diagonal matrix $D$ and $\alpha,~\beta$ run over the flavor indices $e,~\mu,~\tau$. This condition takes the form of a mass sum rule, similar to sec.~\ref{sec:sumrules}, where the coefficients are the mixing matrix elements. We show explicitly the expressions for vanishing mass matrix elements in appendix \ref{sec:massmatrix_el}.
In the case of one-texture zero mass matrices, all six possible matrices are in agreement with experimental data \cite{Lashin:2011dn}, although in some case only one mass ordering is allowed, see tabs.~\ref{tab:mbb_1t0}, \ref{tab:mbb_2t0}.

\begin{table}
\centering
\caption{The fraction of each model that is in the funnel for the 1-texture zero cases as defined in the text using a log prior assuming the NO.
All six models are viable in some region of parameter space.}
\begin{tabular}{c|c}
& Fraction in funnel\\\hline
$M_{ee}$&1\\
$M_{e\mu}$&0.31\\
$M_{e\tau}$&0.30\\
$M_{\mu\mu}$&0\\
$M_{\mu\tau}$&0\\
$M_{\tau\tau}$&0\\
\end{tabular}
\label{tab:mbb_1t0}
\end{table}

\begin{table}
\centering
\caption{The fraction of each model in the funnel for the 2-texture zeros cases as defined in the text assuming the NO.
Models with an X are not viable anywhere in parameter space at $3\sigma$.
}
\begin{tabular}{c|c|c|c|c|c}
& $M_{e\mu}$&$M_{e\tau}$&$M_{\mu\mu}$&$M_{\mu\tau}$&$M_{\tau\tau}$\\\hline
 $M_{ee}$&1&1&X&X&X\\
 $M_{e\mu}$&&X&0&X&0\\
 $M_{e\tau}$&&&0&X&0\\
 $M_{\mu\mu}$&&&&X&0\\
 $M_{\mu\tau}$&&&&&X\\
\end{tabular}
\label{tab:mbb_2t0}
\end{table}

\begin{figure*}
\centering
\includegraphics[width=0.49\textwidth]{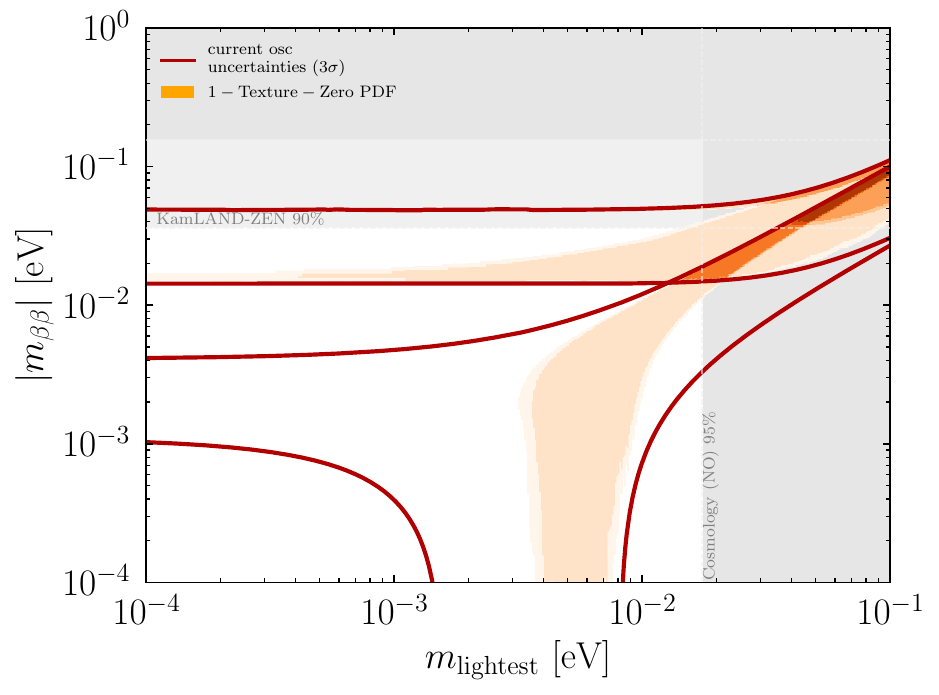}
\includegraphics[width=0.49\textwidth]{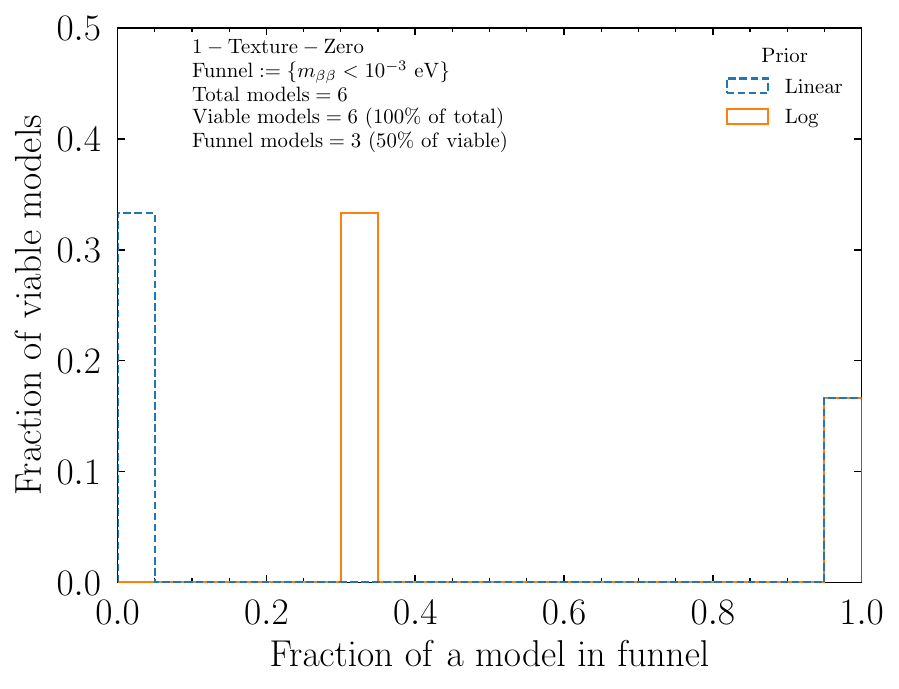}
\caption{The same as fig.~\ref{fig:gCP results} but for 1 texture zeros.
Note that the $M_{ee}=0$ model predicts that $|m_{\beta\beta}|=0$ and is thus at the bottom of the left panel, hence the presence of a model that predicts 100\% of the model space in the funnel.}
\label{fig:1T0 results}
\end{figure*}

There are, in total, 15 two-texture zero matrices of which seven are in agreement with experimental data \cite{Frampton:2002yf,Guo:2002ei,Xing:2002ta,Desai:2002sz,Xing:2002ap,Dev:2006if,Dev:2006xu, Fritzsch:2011qv,Honda:2003pg}.
Two of them feature a vanishing $e-e$ mass matrix element ($M_{ee}=0,~M_{e\mu}=0$ or $M_{ee}=0,~M_{e\tau}=0$) and therefore predict 100\% of the parameter space in the funnel. Upon imposing two vanishing mass matrix elements, we obtain four relations between the mixing matrix elements and observables. 
The two-texture zero mass matrices in agreement with experimental data have one vanishing diagonal element of $M$, $M_{\eta\eta}=0,~\eta=(e,~\mu,~\tau)$, and one of the off-diagonal elements in the electron row vanishes,
$M_{e\gamma}=0$ with $\gamma={\mu,~\tau}$. Lastly, the case with $M_{\mu\mu}=M_{\tau\tau}=0$ is also in agreement with current data. 

\begin{figure*}
\centering
\includegraphics[width=0.49\textwidth]{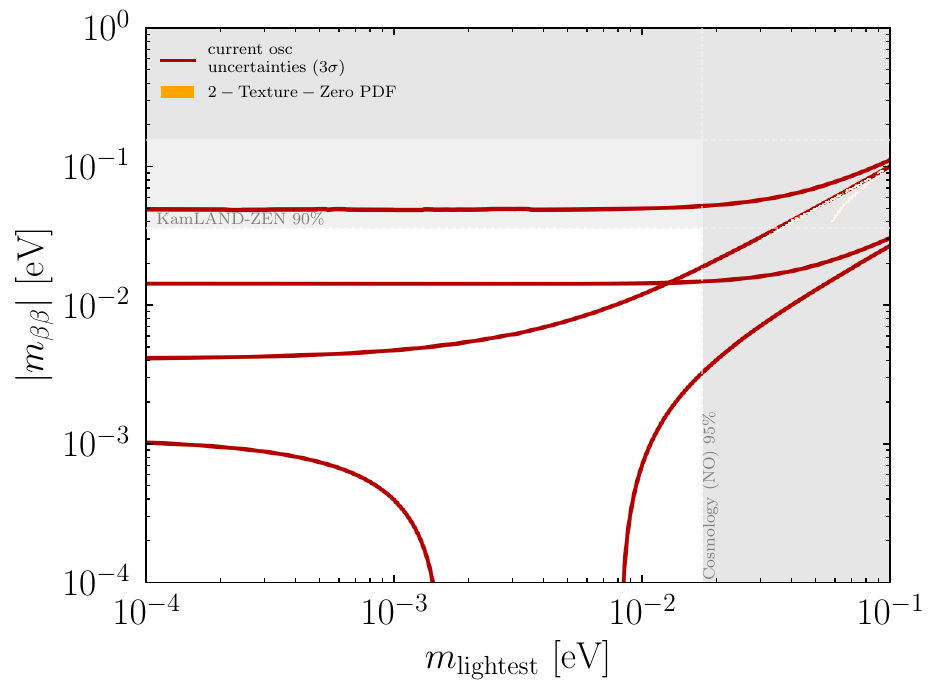}
\includegraphics[width=0.49\textwidth]{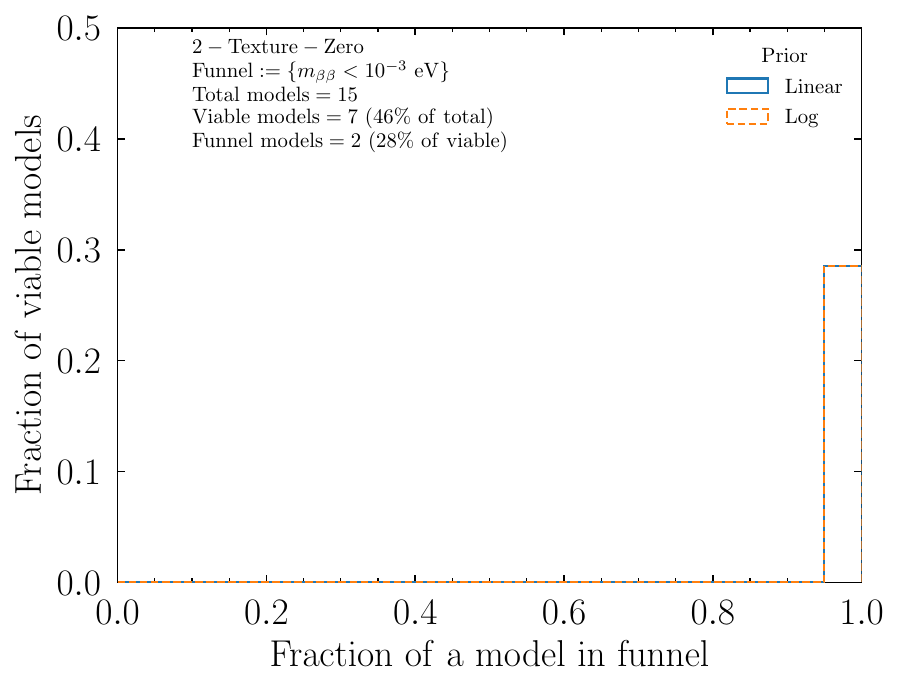}
\caption{The same as fig.~\ref{fig:gCP results} but for 2 texture zeros.
Note that as in fig.~\ref{fig:1T0 results}, the two $M_{ee}=0$ models predict that $|m_{\beta\beta}|=0$ and is thus at the bottom of the left panel, hence the presence of two models that predict 100\% of the model space in the funnel.
On the left panel there is a small sliver of predicted space in the quasi-degenerate region on the top right in the cosmologically disfavored region.}
\label{fig:2T0 results}
\end{figure*}

The one-texture zero case leads to two predictions for $M_\nu$ as it constrains the real and imaginary parts of this mass matrix element\footnote{Even if the mass matrix element is chosen to be real there are still two constraints on the combination of mixing matrix elements and mass eigenvalues.}. Two constraints also apply to oscillation and $0\nu\beta\beta$ experiments. In the two-texture zero case the number vanishing mass matrix elements double; therefore, the number of constraints is now four.
In both one- and two-texture zero cases, one can derive expressions for masses, see appendix \ref{sec:massmatrix_el}.

In fig.~\ref{fig:1T0 results} we show the results for the one-texture zero case. We find that three models are in the funnel where the model with $M_{ee}=0$ is 100\% in the funnel and $M_{e\mu}$ and $M_{e\tau}$ are partially in the funnel. Additionally, both models in the funnel predict a lower bound on $m_1\gtrsim 4\times 10^{-3}$ eV in NO.

In fig.~\ref{fig:2T0 results} we show the results for the two-texture zero case. We find that only $46\%$ (7/15) of the models are viable and 28\% (2/7) of the viable models are in the funnel, specifically the two $M_{ee}=0$ models. Furthermore, the five non-funnel viable models predict large values of the lightest mass $m_\text{lightest}>3\times10^{-2}$ eV, which is in the quasi-degenerate region and is already ruled out by cosmology.
This means the only actual viable models, when also including cosmological data, for two-texture zeros are the models with $M_{ee}=0$ and one of either $M_{e\mu}=0$ or $M_{e\tau}=0$ which also predict $m_{\beta\beta}=0$.
This is a new result. 

There is another study that looked at a unique class of models which can be described as texture zeros with rotational corrections.
This study concluded that it was possible for models to go well into the funnel, although it is important to note that $\theta_{13}$ was not known at the time of this study \cite{Jenkins:2008ex}.

\subsection{Modular symmetries with fixed modulus}
\label{sec:modular}
In \cite{Gehrlein:2020jnr} models based on modular symmetries with a fixed modulus were studied.
In these models only one field is introduced which, upon obtaining a vacuum expectation value, breaks the flavor symmetry \cite{Feruglio:2017spp} (for a review see also \cite{Kobayashi:2023zzc}).
In comparison to models with discrete symmetries where multiple fields are introduced, a reduction of free parameters is achieved which leads to more correlations between physical parameters. So far, five models with the most correlations have been identified in the literature.\footnote{Note that there are models with a free value of the modulus field where a sum rule can arise like in \cite{CentellesChulia:2023zhu}.} In these models the symmetric mixing matrix gets corrected by a 1-2 or 1-3 rotation, similar to the case of one charged lepton rotation. 
Then the three mixing angles and the Dirac CP phase are determined by two free model parameters only.
These models also lead to mass sum rules similar to those discussed in subsection \ref{sec:sumrules}.
In this case, however, the coefficients of the mass sum rule are not constant, but they depend on the two free model parameters, leading to a correlation among the neutrino masses, Majorana phases, and mixing parameters. The expressions for the mass sum rules and the mixing angles can be found in \cite{Gehrlein:2020jnr}; for convenience, we quote them again in appendix \ref{sec:modularsym}.
Similar to the case of mass sum rules these models predict a lower and an upper bound on the lightest mass, see sec.~\ref{sec:sumrules}. 

\begin{figure*}
\centering
\includegraphics[width=0.49\textwidth]{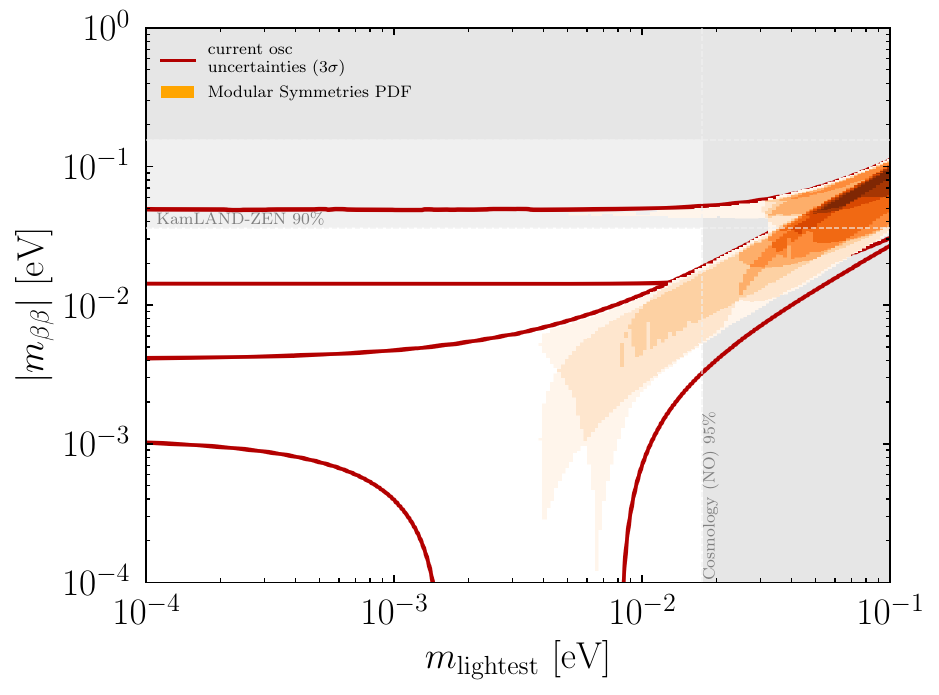}
\includegraphics[width=0.49\textwidth]{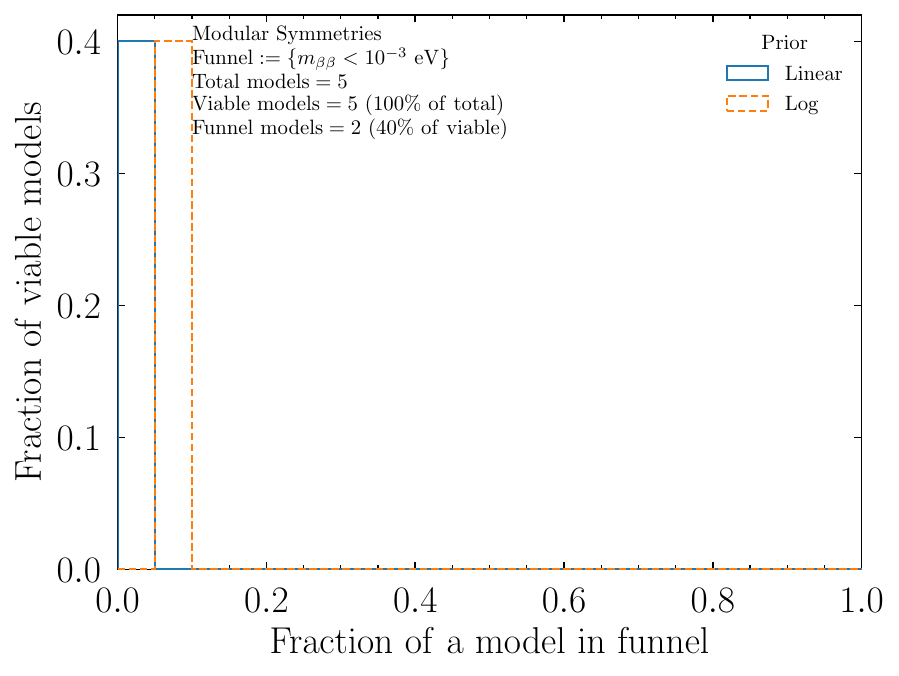}
\caption{The same as fig.~\ref{fig:gCP results} but for models with modular symmetries.}
\label{fig:MOD results}
\end{figure*}

Our results are shown in fig.~\ref{fig:MOD results}.
All five models are viable, although two of them are only valid in the high mass region that is disfavored by cosmological data.
For the five models present in the literature, we find that two are in the funnel at only the 5\% or 7\% level (log prior).

It is likely that more models with such correlations exist.
Their predictivity of different neutrino observables makes them an interesting target for future neutrino experiments, even beyond neutrinoless double beta decay \cite{Gehrlein:2022nss}.

\section{Discussion}
\label{sec:discussion}
In table \ref{tab:mbb_results} we give an overview of the number of models contained in each category and the number of allowed models, and we provide the fraction of models in the funnel.
We see that some model categories have a sizable fraction in the funnel; however, we caution the reader that this could be an over representation due to the measure chosen, see eq.~\eqref{eq:funnel fraction}.
Finally, some model categories only feature a small number of viable models; therefore, the total number of models in the funnel is not so big.

Among the model categories surveyed we find that funnel fractions of $\approx$ 20--100\% are possible, making probes of the funnel region a crucial target to comprehensively test different flavor models.
Interestingly, most of the models studied that feature a fraction of parameter space in the funnel also predict parameter space outside of the funnel. This allows $0\nu\beta\beta$ experiments to narrow down the parameter spaces of the models in the near future, even without penetrating the funnel region. Exceptions to this are texture zero models with $M_{ee}=0$ which predict $m_{\beta\beta}=0$ exactly and are therefore fully contained in the funnel\footnote{Potentially, some sum rules with coefficients $c_i>2$ or different values of $d$ might also be fully contained in the funnel.}.

\begin{table}
\centering
\caption{An overview of the number of total model groups contained in each model category, the number of valid model groups given oscillation data, and the fraction of valid model groups which penetrate the funnel region in the NO.}
\begin{tabular}{l|c|c|c}
&&&Fraction of viable\\
Model&Total&Viable & models in funnel\\\hline
Generalized CP &10& 10 &0.20\\
Mass sum rules &3137&1968 & 0.14\\
Charged lepton corrections &75& 31 &1.00\\
1-Texture zeros &6& 6&0.50\\
2-Texture zeros &15&7&0.28\\
Modular symmetries &5 & 5 & 0.40
\end{tabular}
\label{tab:mbb_results}
\end{table}

Here we focused on models based on symmetries. Another approach, referred to as ``anarchy" assumes that the leptonic mixing matrix can be described as the result of a random draw from an unbiased distribution of unitary three-by-three matrices \cite{Hall:1999sn,Haba:2000be,deGouvea:2012ac}. In the past it has been shown that the probability for $|m_{\beta\beta}|<10^{-3}$ eV is small, around $5\%$ \cite{deGouvea:2012ac}, see also \cite{Jenkins:2008ms}.
Therefore, flavor models based on symmetries can be more likely to predict a region in the funnel in some cases.

Additionally, several models like sum rules, modular symmetries, and 1-texture-zeros prefer large values for the lightest mass and present a lower limit on $m_\text{lightest}$ within the reach of near future 0$\nu\beta\beta$ experiments \cite{Agostini:2022zub} such that the whole region of parameter space can be probed with cosmology very soon. 
Generalized CP and charged lepton corrections, on the other hand, do not predict the absolute mass scale such that these models will remain viable independent of a future measurement of $m_\text{lightest}$. 
We note that these models also predict other observables which allow us to test these models.
In fact, these predictions are crucial in our assessment of the validity of these models. We find that for charged lepton corrections, and 2-texture zeros only roughly half of the total models are viable due to their predictions for the mixing angles. 
In comparison to previous studies in the literature, an important change is that the precision on $\Delta m_{31}^2$ has improved from 4\% to 1\% \cite{Denton:2022een}, which has a big effect on the results and, in particular, the validity of models.
Future measurements of the oscillation parameters will further test models; in particular, improvements on $\delta$ and $\theta_{12}$ will probe and distinguish different models \cite{Gehrlein:2022nss} which will further narrow down the number of valid flavor models. Even though neither $\delta$ nor $\theta_{23}$, the two oscillation parameters which are currently most uncertain, plays a role for $0\nu\beta\beta$, their measurements indirectly affect our results here as these measurements constrain the model parameters in that they tell us which models are valid.
For this study we did not include a prior on $\delta$; however, models based on generalized CP, charged lepton corrections, texture zeros, and models with modular symmetries also predict $\delta$; their validity will be tested with the next generation of experiments, which in turn will change the 0$\nu\beta\beta$ landscape.
Therefore, flavor models provide a rich model space to test with upcoming experiments, including oscillation experiments and cosmology.

When quantifying the fraction of a model in a funnel, some choices need to be made related to eq.~\ref{eq:funnel fraction}.
While many of our results are presented with a log prior in $m_{\rm lightest}$, we also perform all the same calculations with a linear prior in $m_{\rm lightest}$; see the right panel in figs.~\ref{fig:gCP results}, \ref{fig:SR results}, \ref{fig:CLC results}, \ref{fig:1T0 results}, \ref{fig:2T0 results}, and \ref{fig:MOD results}.
A linear prior in masses is related to what one would expect in anarchy for certain mass models \cite{Haba:2000be}\footnote{For real Dirac matrices the mass part of the anarchy measure is $\Delta m^2_{21}\Delta m^2_{31}\Delta m^2_{32}dm_1dm_2dm_2$ while the complex cases have an additional weighting of $m_1m_2m_3$ which modifies the prior.} while, on the other hand, the other fermion masses (i.e.~their Higgs Yukawa couplings) seem to be distributed uniformly on a log scale, see e.g.~\cite{Donoghue:2009me}.
When considering a linear prior, the fraction of the model prediction in the funnel is less than or equal to the fraction in the funnel with a log prior.
They are the same in models which are entirely in the funnel (such as the $M_{ee}=0$ texture zero model predictions) or not at all in the funnel.
The difference is because models that are partially in the funnel all contain predictions out of the funnel at higher values of $m_{\rm lightest}$.
With a linear prior, this region tends to quickly dominate the comparatively small region in the funnel.
Note that it is conceivable that a model prediction would have the opposite trend: a larger funnel fraction with a linear prior than with a log prior.
However, that does not seem to happen with the model predictions considered here.

Finally, in all of the above computations, we assume that the predictions in the models are exact at the low energy scale of neutrinoless double beta decay, and we assume that the full particle content of the mass model of neutrinos does not modify these symmetry predictions.
In models where the symmetry arises from a high scale, presumably related to neutrino mass generation, potential corrections to the low scale predictions can arise, however, from renormalization group running effects.
In this case there may be corrections from running that could have an important impact on the parameters depending on both the flavor prediction structure and the mass generation model, see e.g.~\cite{Antusch:2003kp,Antusch:2005gp,Schmidt:2007nq,Chakrabortty:2008zh,Ohlsson:2013xva,Merle:2015ica,Gehrlein:2015ena,Gehrlein:2016wlc}.
In some cases such as texture zeros, the flavor prediction may persist down to low scales \cite{Dev:2014dla,Fritzsch:2011qv} however, and thus running has no impact in these cases.
On the other hand, it has been shown that in some mass generation models the neutrino parameters do not run from the heavy scale to the low scale at all \cite{Bergstrom:2010qb}; thus, any flavor prediction associated with the mass generation mechanism would be preserved at the scale of oscillations and neutrinoless double beta decay.
In addition, in the case of random values (e.g.~anarchy) of the mixing parameters \cite{Hall:1999sn,deGouvea:2012ac}, running effects do not significantly change the funnel narrative under various mass generation scenarios \cite{Brdar:2015jwo}.
Fully exploring this space of both flavor predictions and underlying mass models at the same time is beyond the scope of this paper.
We can estimate that, in these scenarios the flavor predictions of, for example, all of the sum rules considered that land in the funnel may not be the ones we have shown; however, we anticipate that a comparable fraction of them will be in the funnel when scanning over many different mass generation scenarios, and thus our qualitative results should be independent of renormalization group running.

\section{Conclusions}
\label{sec:conclusions}
An observation of neutrinoless double beta decay will have a tremendous impact on our understanding of Nature. Apart from proving that lepton number is not a conserved symmetry of Nature, it can also provide valuable insights into other open problems of the SM like the flavor puzzle.
Motivated by the current and anticipated experimental progress of various neutrino experiments, we have studied the predicted ranges of $m_{\beta\beta}$ and $m_\text{lightest}$ of several classes of flavor models. In particular, we focused on the funnel region in normal mass ordering with $|m_{\beta\beta}|<1$ meV, which is experimentally challenging to probe in order to answer the question of how likely it is that a model prediction is only realized in the funnel, which would require a massive leap in experimental progress.

We have considered five broad classes of flavor models based on different symmetries. After assessing their validity by comparing their predictions to our up-to-date experimental knowledge from oscillation experiments, we calculated the funnel fractions of the valid models.
Our study shows that all of the studied model classes feature models with parameter space in the funnel. Indeed, the fractions of viable models that are in the funnel range from 5--100\%.
Thus, flavor models may well be more likely to predict that $|m_{\beta\beta}|$ is in the funnel than in the case of random neutrino mixing matrices, anarchy, where the funnel probability is around 5\%.
Additionally, we have provided PDFs of the predicted $m_{\beta\beta}-m_\text{lightest}$ regions of the classes of flavor models.
We find that models which predict the absolute mass scale generally predict larger neutrino masses such that cosmological observatories can test them as well in the near future in addition to crucial tests of the predicted values for the mixing angles to upcoming oscillation experiments.

Our results can be used to plan the target sensitivity of upcoming neutrinoless double beta decay experiments with the goal to probe most of the parameter space motivated by flavor models
(see \cite{Agostini:2015dna} for a similar study focusing on existing sum rules in flavor models).

In this study we focused on light Majorana neutrino exchange as the underlying scenario for $0\nu\beta\beta$. 
Other scenarios when new particles are introduced could also predict a region in the funnel. For example, models with a sterile neutrino allow for vanishing rates for 0$\nu\beta\beta$ \cite{Abada:2018qok,Zhao:2022trb,Asaka:2020wfo,Asaka:2020lsx,Asaka:2021hkg,Pascoli:2013fiz}. In particular, depending on the sterile parameters, a funnel in IO opens up.
However, a Bayesian analysis of eV sterile parameters showed that the posterior probability that $|m_{ee}|$ falls into the funnel region is very small, $<0.3\%$ \cite{Huang:2019qvq}. On the other hand, models based on a left-right symmetry do not predict a region in the funnel \cite{Tello:2010am,Li:2020flq}, neither does a model where a new scalar interaction \cite{Graf:2020cbf} is introduced. 

As neutrinoless double beta decay experiments continue to push the limits down into the inverted mass ordering region, understanding the theoretically favored regions of parameter space is important to plan for experimental upgrades.
In order to unambiguously interpret a measurement in the context of flavor models presented in this work, improvements in nuclear matrix calculations are also needed.

\begin{acknowledgments}
The authors acknowledge support by the United States Department of Energy under Grant Contract No.~DE-SC0012704.
Part of this research was completed while at the Center for Theoretical Underground Physics and Related Areas (CETUP*).
\end{acknowledgments}

\appendix
\section{Expressions for the elements of the mass matrix}
\label{sec:massmatrix_el}
Here we give the expressions for the elements of the mass matrix as a function of the mixing parameters and the mass eigenvalues.
The number of free parameters in the Majorana mass matrix and in the mixing matrix together with the light mass eigenvalues, nine, coincide as expected.
Realistically, we are only able to measure eight out of the nine free parameters in the mass matrix, as we have no observable which depends on the individual Majorana phases.
\begin{widetext}
\begin{align}
m_{ee}={}&m_1\text{e}^{\text{i}\alpha} c_{12}^2 c_{13}^2 +m_2\text{e}^{\text{i}\beta} 
c_{13}^2s_{12}^2+m_3s_{13}^2\\
m_{e\mu}={}&m_1 \text{e}^{\text{i}\alpha} c_{12}c_{13} (-c_{23}s_{12}- \text{e}^{\text{i}\delta} c_{12}s_{13}s_{23}) +m_2 \text{e}^{\text{i}\beta}c_{13}s_{12}(c_{12}c_{23}- \text{e}^{\text{i} \delta} s_{12}s_{13}s_{23})\nonumber\\& +m_3 \text{e}^{\text{i}\delta} c_{13}s_{13}s_{23}\\
m_{e\tau}={}&m_1 \text{e}^{\text{i}\alpha}
c_{12}c_{13}(- c_{12}c_{23}s_{13}+\text{e}^{-\text{i}\delta}s_{12}s_{23}) +m_2 \text{e}^{\text{i}\beta}c_{13}s_{12}(-c_{23} s_{12} s_{13}-\text{e}^{-\text{i}\delta}c_{12}s_{23})\nonumber\\& +m_3 c_{13}c_{23}s_{13}\\
m_{\mu \mu}={}&m_1 \text{e}^{\text{i}\alpha} (-c_{23}s_{12}-\text{e}^{\text{i}\delta}c_{12}s_{13}s_{23})^2
+m_2 \text{e}^{\text{i}\beta} (c_{12}c_{23}-\text{e}^{\text{i}\delta} s_{12}s_{13}s_{23})^2+m_3 \text{e}^{2\text{i}\delta} c_{13}^2s_{23}^2\\
m_{\mu \tau}={}&m_1 \text{e}^{\text{i}\alpha}(-c_{12}c_{23}s_{13}+
\text{e}^{-\text{i}\delta}s_{12}s_{23})(-c_{23}s_{12}-\text{e}^{\text{i}\delta}c_{12}s_{13}s_{23})\nonumber\\&+m_2 \text{e}^{\text{i}\beta}
(-c_{23}s_{12}s_{13}-\text{e}^{-\text{i}\delta}
c_{12}s_{23})(c_{12}c_{23}-\text{e}^{\text{i}\delta}s_{12}s_{13}s_{23}) +m_3\text{e}^{\text{i}\delta}c_{13}^2c_{23}s_{23} \\
m_{\tau \tau}={}&m_1 \text{e}^{\text{i}\alpha} (-c_{12}c_{23}s_{13}+\text{e}^{-\text{i}\delta}s_{12}s_{23})^2+m_2 \text{e}^{\text{i}\beta} (-c_{23}s_{12}s_{13}-\text{e}^{-\text{i}\delta}c_{12}s_{23})^2 \nonumber\\&+m_3 c_{13}^2c_{23}^2 
\end{align}
\end{widetext}

For one vanishing matrix element $M_{\alpha\beta}=0$, the expressions for the neutrino mass ratios are \cite{Lashin:2011dn}
\begin{widetext}
\begin{align}
\frac{m_1}{m_3}&=\frac{\text{Re}(U_{\alpha3}U_{\beta3})\text{Im}(U_{\alpha2}U_{\beta2}\text{e}^{\text{i}\beta})-\text{Re}(U_{\alpha2}U_{\beta2}\text{e}^{\text{i}\beta})\text{Im}(U_{\alpha3}U_{\beta3})}{\text{Re}(U_{\alpha2}U_{\beta2}\text{e}^{\text{i}\beta})\text{Im}(U_{\alpha1}U_{\beta1}\text{e}^{\text{i}\alpha})- \text{Im}(U_{\alpha2}U_{\beta2}\text{e}^{\text{i}\beta})\text{Re}(U_{\alpha1}U_{\beta1}\text{e}^{\text{i}\alpha})}\\
\frac{m_2}{m_3}&=\frac{\text{Re}(U_{\alpha1}U_{\beta1}\text{e}^{\text{i}\alpha})\text{Im}(U_{\alpha3}U_{\beta3}\text{e}^{\text{i}\beta})-\text{Im}(U_{\alpha1}U_{\beta1}\text{e}^{\text{i}\alpha})\text{Re}(U_{\alpha3}U_{\beta3}\text{e}^{\text{i}\beta})}{\text{Re}(U_{\alpha2}U_{\beta2}\text{e}^{\text{i}\beta})\text{Im}(U_{\alpha1}U_{\beta1}\text{e}^{\text{i}\alpha})- \text{Im}(U_{\alpha2}U_{\beta2}\text{e}^{\text{i}\beta})\text{Re}(U_{\alpha1}U_{\beta1}\text{e}^{\text{i}\alpha})}
\end{align}
\end{widetext}

The condition of two vanishing mass matrix elements $M_{\alpha\beta},~M_{\delta\gamma}$, $(\alpha\beta)\neq (\delta\gamma)$ can be translated to expressions for the neutrino masses and Majorana phases
\cite{Xing:2002ta}
\begin{align}
\frac{m_1}{m_3} & = \left |
\frac{U_{\gamma3} U_{\delta3} U_{\alpha 2} U_{\beta 2} - U_{\gamma 2} U_{\delta2} U_{\alpha 3}
U_{\beta 3}}{U_{\gamma 2} U_{\delta 2} U_{\alpha 1} U_{\beta 1} - U_{\gamma 1} U_{\delta 1}
U_{\alpha 2} U_{\beta 2}} \right | \, ,
\nonumber\\ \nonumber \\
\frac{m_2}{m_3} & = \left |
\frac{U_{\gamma 1} U_{\delta1} U_{\alpha 3} U_{\beta 3} - U_{\gamma 3} U_{\delta 3} U_{\alpha 1}
U_{\beta 1}}{U_{\gamma 2} U_{\delta2} U_{\alpha 1} U_{\beta 1} - U_{\gamma 1} U_{\delta 1}
U_{\alpha 2} U_{\beta 2}} \right | \, .
\label{eq:massratioSR}
\end{align}

\begin{align}
\alpha & =
 \arg \left [ \frac{U_{\gamma 3} U_{\delta 3} U_{\alpha 2} U_{\beta 2}
- U_{\gamma 2} U_{\delta 2} U_{\alpha 3}
U_{\beta 3}}{U_{\gamma2} U_{\delta 2} U_{\alpha 1} U_{\beta 1} - U_{\gamma 1} U_{\delta 1}
U_{\alpha 2} U_{\beta 2}} \right ] \, ,
\nonumber \\ \nonumber \\
\beta & = \arg \left [
\frac{U_{\gamma 1} U_{\delta 1} U_{\alpha 3} U_{\beta 3} - U_{\gamma 3} U_{\delta 3} U_{\alpha 1}
U_{\beta 1}}{U_{\gamma 2} U_{\delta 2} U_{\alpha 1} U_{\beta 1} - U_{\gamma 1} U_{\delta 1}
U_{\alpha 2} U_{\beta 2}} \right ] \, .
\end{align}
We see that the Majorana phases depend on the value of Dirac CP phase in the PMNS matrix contained in matrix elements $U_{\mu i}$, $i\in[1,3]$, $U_{\tau 1},~U_{\tau 2}$.
Furthermore, from eq.~\eqref{eq:massratioSR} we see that the ratios of the neutrino masses depend on the values of the matrix elements. With the known values for the mass splittings, we obtain a lower bound on the lightest mass, depending on which matrix elements are zero.

From these expressions we see that there is no one-to-one correspondence between mass eigenvalues and observables in experiments. This means that only with a combination of measurements (i.e.~different oscillation channels and an observation of neutrinoless double beta decay) one can reconstruct the neutrino mass matrix. 
This situation is similar to considering only one measurement at oscillation experiments. In one channel one is only sensitive to a certain combination of parameters. Only a combination of measurements can tell us the values of the mixing angles.

In addition, for absolute neutrino mass measurements such as from KATRIN or cosmology and neutrinoless double beta decay we are left with one measurement of one combination of parameters  assuming no prior knowledge of the results from other experiments; one can therefore predict something for oscillation experiments as well.
In reality, we already have measurements from oscillations such that predictions from an absolute mass measurement do not contribute new knowledge for oscillation experiments.
\section{Gell-Man \texorpdfstring{$SU(3)$}{SU(3)} generators and the mass matrix}
\label{sec:gell-man}
The mass matrix need not be parameterized as three masses, three mixing angles, and three phases.
Other parameterizations are possible.
One such explicit example is with $SU(3)$ generators, such as the Gell-Mann matrices, see e.g.~\cite{Krivoruchenko:2023npj}.
That is, the mass matrix from eq.~\eqref{eq:diag} can be written as
\begin{equation}
M=M_{\rm scale}\prod_{i=1}^8\exp(a_i\lambda_i)\,,
\end{equation}
where the eight $a_i\in\mathbb R$ are free parameters as is $M_{\rm scale}$ which sets the dimensionful scale and the $\lambda_i$ are some traceless representation of $SU(3)$ such as the Gell-Mann matrices.
The dimensionful scale parameter can also be thought of as the trace part of $M$.
This could imply novel flavor structures similar to texture zeros by requiring some subset of the $a_i$ to be zero.
One could also consider representations other than the Gell-Mann matrices, such as cyclic representations \cite{Harrison:2014epa}.
Investigating the phenomenology of such flavor models is beyond the scope of this work.

\section{Independent generalized CP models}
\label{sec:independentgCP}
In table \ref{tab:genCPmodels} we list the phase combinations which are independent for models with generalized CP.
We see that it is sufficient to constrain $\alpha$ to be between $[0,\pi]$ and $\beta\in [0,2\pi]$ to cover the whole parameter space.
\begin{table}
\centering
\caption{Pairs of values for the Majorana phases $\alpha,~\beta$ in models with generalized CP which lead to different results for $|m_{\beta\beta}|$. Some pairs are equivalent to others; these are in the table to the right. The bolded pairs are the ones which predict a region in the funnel.}
\begin{tabular}{c}
$(\alpha,~\beta$) \\\hline
$\mathbf{(0,\boldsymbol{\pi})}$\\
$\mathbf{(\boldsymbol{\pi},0)}$\\
$(0,0)$\\
$(\pi,\pi)$\\
\end{tabular}\hspace{3cm}
\begin{tabular}{c}
$(\alpha,~\beta$) \\\hline
$(0,\pi/2)$ or $(0,3\pi/2)$\\
$(\pi/2,3\pi/2)$ or $(3\pi/2,\pi/2)$\\
$(\pi,\pi/2)$ or $(\pi,3\pi/2)$\\
$(\pi/2,0)$ or $(3\pi/2,0)$\\
$(\pi/2,\pi/2)$ or $(3\pi/2,3\pi/2)$\\ $(\pi/2,\pi)$ or $(3\pi/2,\pi)$\\
\end{tabular}
\label{tab:genCPmodels}
\end{table}

\section{Sum rules in the funnel}
\label{sec:msrfunnel}
In table \ref{tab:SRmodelsfunnel} we show the parameters of sum rules which lead to at least a $50\%$ fraction in the funnel with a log prior.

\begin{table}
\centering
\caption{Parameters of sum rules which lead to at least a 50\% fraction in the funnel with a log prior.}
\begin{tabular}{c|c|c|c|c|c}
$c_1$&$c_2$&$d$&$\chi_1$&$\chi_2$ &Fraction in funnel\\\hline
1&2&$-1/2$&$\pi/2$&0&0.74\\
1&2&$-1/2$&$3\pi/2$&0&0.74\\
4/6&1&$-1/2$&$3\pi/2$&0&0.67\\
4/6&1&$-1/2$&$\pi/2$&0&0.62\\
5/6&1&$-1/2$&$3\pi/2$&0&0.59\\
5/6&1&$-1/2$&$\pi/2$&0&0.58\\
5/6&2&$-1/2$&$\pi/2$&0&0.58\\
5/6&2&$-1/2$&$3\pi/2$&0&0.58\\
1&2&$-1/2$&0&$\pi/2$&0.58\\
1&2&$-1/2$&0&$3\pi/2$&0.58\\
1/3&1&$-1/2$&0&$\pi/2$&0.56\\
 4/6&5/6&$-1/2$&$\pi/2$&0&0.54\\
4/6&5/6&$-1/2$&$3\pi/2$&0&0.54\\
1/6&1/6&$-1/2$&$\pi$&$\pi$&0.54\\
1/3&1&$-1/2$&0&$3\pi/2$&0.54\\
1/6&1/2&$-1$&0&0&0.51\\
 1/6& 4/6&$-1$&0&0&0.51\\
\end{tabular}
\label{tab:SRmodelsfunnel}
\end{table}

\section{Expressions for physical parameters in models with modular symmetries}
\label{sec:modularsym}
Here we give the expressions for the oscillation parameters and the sum rule in models with modular symmetries, first derived in \cite{Gehrlein:2020jnr}. The parameters $\theta,~\phi$ are free model parameters.
\begin{itemize}
\item A model based on $A_4$ symmetry was studied in \cite{Novichkov:2018yse}. Two cases arise, depending on the assumption on the charged lepton mixing matrix. The expressions for the mixing parameters remain the same in both scenarios.
\begin{align}
\label{eq:sr1_mix1}
 \sin^2 \theta_{12}(\theta) &= \frac{1}{3 - 2 \sin^2 \theta} \,, \\
 \label{eq:sr1_mix2}
 \sin^2 \theta_{13}(\theta) &= \frac{2}{3} \sin^2 \theta \,, \\
 \label{eq:sr1_mix3}
 \sin^2 \theta_{23}(\theta,\phi) &= \frac{1}{2} + \frac{\sin \theta_{13}(\theta) }{2} \frac{\sqrt{2 - 3 \sin^2 \theta_{13}(\theta) }}{1 - \sin^2 \theta_{13}(\theta) } \cos \phi \,, \\
 \label{eq:sr1_mix4}
 \delta(\theta,\phi) &= \arcsin\left(- \frac{\sin \phi}{\sin 2 \theta_{23}(\theta,\phi) } \right) \,,
\end{align}
The parameters in the sum rule (see eq.~\eqref{eq:msr}) in case I are
\begin{align}
 c_1 &= - \text{e}^{-2 \ci \phi} - \ci \text{e}^{-\ci \phi} f_2 \sin \phi \,, \nonumber\\
 &= - \text{e}^{-2 \ci \phi} \frac{\sqrt{3} \sin(2\theta) - \cos \phi \cos(2\theta) - \ci \sin \phi}{\sqrt{3} \sin(2\theta) - \cos \phi \cos(2\theta) + \ci \sin \phi} \\
 c_2 &= - \text{e}^{-\ci \phi} \frac{ 2 }{ \sqrt{3} \sin(2\theta) - \cos \phi \cos(2\theta) + \ci \sin \phi } \,\\
 d&=1
\end{align}
In the second scenario the coefficients of the mass sum rule are related to the coefficients in the first case by
\begin{align}
c_1^{(\text{II})} &= c_1^{(\text{I})} \text{e}^{-4 \ci \phi} \,, \\ 
c_2^{(\text{II})} &= - c_2^{(\text{I})} \text{e}^{2 \ci \phi } \,.
\end{align}

\item A model based on two modular $S_4$ symmetries has been studied in \cite{King:2019vhv}:
\begin{align}
\sin\theta_{13}&=\frac{\sin\theta}{\sqrt{3}}~,\\
\tan\theta_{12}&= \frac{\cos\theta}{\sqrt{2}}~,\\
\tan \theta_{23}& =\left|\frac{\cos\theta+\sqrt{\frac{2}{3}}\text{e}^{\ci \phi} \sin\theta}{\cos\theta-\sqrt{\frac{2}{3}}\text{e}^{\ci \phi}\sin\theta}\right|~.\\
 \tan \delta& = - \frac{ 5 + \cos(2 \theta)}{1+ 5 \cos (2 \theta) } \tan\phi \,.
\end{align}
The parameters in the sum rule read
\begin{align}
c_1 &= \frac{1}{\cos^2 \theta - \text{e}^{\ci \phi} \sin (2 \theta)} \,, \\
c_2 &= -\frac{\tan \theta + 2 \, \text{e}^{\ci \phi}}{2 \, \text{e}^{3 \ci \phi} - \text{e}^{2 \ci \phi} \cot (\theta)} \,, \\
d &= -1 \,.
\end{align}

\item In \cite{Novichkov:2018ovf} a model with a modular $S_4$ symmetry has been investigated:
\begin{align}
\sin \theta_{13} &= \frac{1}{\sqrt{3}} \sin \theta \,,\\
\tan \theta_{12} &= \frac{1}{\sqrt{2}} \cos \theta \,,\\ 
\tan \theta_{23} &= \left| \frac{2 \, \text{e}^{\ci \phi} \tan \theta + \sqrt{3/2} \left(1+\ci \sqrt{3}\right) }
{3\sqrt{2/3} - \left(1-\sqrt{3} \ci \right) \text{e}^{\ci \phi} \tan \theta}\right| \,,\\ 
\tan \delta &= - \frac{(\cos (2 \theta)+5) \left(\sqrt{3} \sin \phi - 3 \cos\phi \right)}{(5 \cos (2 \theta)+1) \left(\sqrt{3} \cos \phi+ 3 \sin \phi \right)} \,.
\end{align}
The parameters of the sum rule are
\begin{align}
f_1&=\frac{2/(\cos \theta \sin \theta)}{(-2 - 2 \ci \sqrt{3}) \text{e}^{\ci \phi} + 
\ci (\ci + \sqrt{3})\cot\theta} 
 ~,\\
f_2&= -\frac{(\ci + \sqrt{3} + 2 (-\ci + \sqrt{3}) \text{e}^{\ci \phi} \cot\theta) \tan\theta}{
 2 (-\ci +\sqrt{3}) \text{e}^{3 \ci \phi} - (\ci + \sqrt{3}) \text{e}^{2 \ci \phi} \cot\theta} 
 ~,\\
d&=-1 \,.
\end{align}
\item The model studied in \cite{Novichkov:2018nkm} is based on a $A_5$ symmetry which leads to the following expressions with $\phi_g=(1 + \sqrt{5})/2$ the golden ratio:
\begin{align}
\sin \theta_{13} ={}& \sqrt{ \frac{1}{10} (5 + \sqrt{5}) } \sin \theta \,,\\
\tan \theta_{12} ={}& \frac{2}{1 + \sqrt{5} } \frac{1}{\cos \theta} \,,\\ 
\tan \theta_{23} ={}& \left| \frac{\sqrt{ \sqrt{5}\phi_g} - \text{e}^{-\ci \phi} \tan \theta}{\sqrt{ \sqrt{5} \phi_g} + \text{e}^{-\ci \phi} \tan \theta }\right| \,, \\
\tan \delta ={}& \frac{4 \sqrt{5+\sqrt{5}} \sin (\phi) \left(2 \left( \sqrt{5}+2\right) \cos ^2(\theta) +1+\sqrt{5}\right)}
{D_\delta} \,, \\
D_\delta ={}& 4 \sqrt{5+\sqrt{5}} \cos (\phi) \cos (2 \theta) \left[ (\sqrt{5}+2) \cos (2\theta)\right.\nonumber\\
&\left.+3+2\sqrt{5} \right]+\sqrt{2} \sin (2\theta) \left[ (5\sqrt{5}+11) \cos (2 \theta)\right.\nonumber\\
&\left.+19+9 \sqrt{5}\right] \cos (2\theta_{23}) \,.
\end{align}
The coefficients of the sum rule are
\begin{align}
c_1 &= \text{e}^{-2 \ci \phi} \frac{\left(1-\sqrt{5}\right) \text{e}^{2 \ci \phi} \cot \theta+\left(\sqrt{5}+1\right) \tan
\theta -8 \text{e}^{\ci \phi}}{\left(1-\sqrt{5}\right) \text{e}^{2 \ci \phi} \tan \theta+\left(\sqrt{5}+1\right) \cot
\theta+8 \text{e}^{ \ci \phi}} \,, \\
c_2 &= \frac{10}{\left(\sqrt{5}-5\right) \text{e}^{2 \ci \phi} \sin^2 \theta + 4\sqrt{5} \text{e}^{\ci \phi} \sin(2 \theta)+\left(5+\sqrt{5}\right) \cos^2 \theta} \,,\\
d&= 1 \,.
\end{align}
\end{itemize}

\bibliography{main}

\end{document}